\documentclass[twocolumn,superscriptaddress,floatfix,preprintnumbers,prl ,nofootinbib,hyperref]{revtex4-2}
\usepackage[colorlinks=true,breaklinks=true]{hyperref}
\usepackage{comment}
\usepackage[T5,T1]{fontenc}

\usepackage[utf8]{inputenc}
\hypersetup{allcolors=[rgb]{0.0 0.0 0.6},linkcolor=[rgb]{0.75 0.05 0.05}}
\usepackage{mathtools}
\usepackage[dvipsnames]{xcolor}
\usepackage{float}
\usepackage{letltxmacro}
\LetLtxMacro{\oldcite}{\cite}
\renewcommand{\cite}[1]{\mbox{\oldcite{#1}}}

\usepackage{orcidlink}

\long\def\exclude#1{}

\bibpunct{[}{]}{,}{n}{}{,}

\begin{document}

\title{Strong Supernova 1987A Constraints on Bosons Decaying to Neutrinos}

\author{Damiano F.\ G.\ Fiorillo \orcidlink{0000-0003-4927-9850}} 
\affiliation{Niels Bohr International Academy, Niels Bohr Institute,
University of Copenhagen, 2100 Copenhagen, Denmark}

\author{Georg G.\ Raffelt
\orcidlink{0000-0002-0199-9560}}
\affiliation{Max-Planck-Institut f\"ur Physik (Werner-Heisenberg-Institut), F\"ohringer Ring 6, 80805 M\"unchen, Germany}

\author{Edoardo Vitagliano
\orcidlink{0000-0001-7847-1281}}
\affiliation{Department of Physics and Astronomy, University of California, Los Angeles, California 90095-1547, USA}

\date{September 23, 2022}

\begin{abstract}

Majoron-like bosons would emerge from a supernova (SN) core by neutrino coalescence of the form $\nu\nu\to\phi$ and $\bar\nu\bar\nu\to\phi$ with 100 MeV-range energies. Subsequent decays to (anti)neutrinos of all flavors provide a flux component with energies much larger than the usual flux from the ``neutrino sphere.'' The absence of 100 MeV-range events in the Kamiokande-II and Irvine-Michigan-Brookhaven signal of SN~1987A implies that less than 1\% of the total energy was thus emitted and provides the strongest constraint on the Majoron-neutrino coupling of $g\alt 10^{-9}\,{\rm MeV}/m_\phi$ for $100~{\rm eV}\alt m_\phi\alt100~{\rm MeV}$. It is straightforward to extend our new argument to other hypothetical feebly interacting particles.

\end{abstract}

\maketitle

{\bf\textit{Introduction.}}---The hot, dense cores of collapsing stars are powerful test beds for novel feebly interacting particles (FIPs), such as sterile neutrinos, dark photons, new scalars, axions and axion-like particles, and many others~\cite{Raffelt:1996wa, Raffelt:2006cw, DiLuzio:2021ysg}, notably including ``secret'' neutrino-neutrino interactions \cite{Heurtier:2016otg, Brune:2018sab, Berryman:2022hds, Akita:2022etk, Chang:2022aas}. In standard SN theory, the trapped electron-lepton number (some 0.30 per baryon) and the gravitational binding energy (some 10\% of the formed neutron star's mass) are carried away by neutrinos on a time scale of a few seconds. The neutrino burst from the historical SN 1987A was observed in the Kamiokande-II \cite{Kamiokande-II:1987idp, Hirata:1988ad, Hirata:1991td, Koshiba:1992yb, Oyama:2021oqp} and Irvine–Michigan–Brookhaven (IMB) \cite{Bionta:1987qt, IMB:1988suc, 1987svoboda} water Cherenkov detectors and the Baksan Underground Scintillation Telescope (BUST) \cite{Alekseev:1987ej, Alekseev:1988gp}. Despite sparse statistics and several anomalies, it has been taken to confirm the standard picture, leaving only limited room for energy loss in the form of FIPs.

If the FIPs interact so strongly that they are trapped themselves or decay before leaving the SN, they contribute to energy transfer \cite{Caputo:2022rca} and may strongly affect overall SN physics and the explosion mechanism. A class of low-explosion-energy SNe provides particularly strong constraints on such scenarios \cite{Caputo:2022mah}. FIPs on the trapping side of the SN-excluded regime are often constrained by other arguments, although allowed gaps may remain, such as the historical ``hadronic axion window'' or more recently the ``cosmic triangle'' for axion-like particles, both meanwhile closed. 

Radiative decays en route to Earth and beyond provide strong limits using $\gamma$-ray observations from SN~1987A and the cosmic diffuse background \cite{Giannotti:2010ty, Oberauer:1993yr, Jaeckel:2017tud, Caputo:2021rux, Calore:2020tjw, Ferreira:2022xlw}. Similar arguments pertain to kilonovae \cite{Diamond:2021ekg} and hypernovae \cite{Caputo:2021kcv}.

\begin{figure}[b!]
\vskip-12pt
    \centering
    \includegraphics[width=0.95\columnwidth]{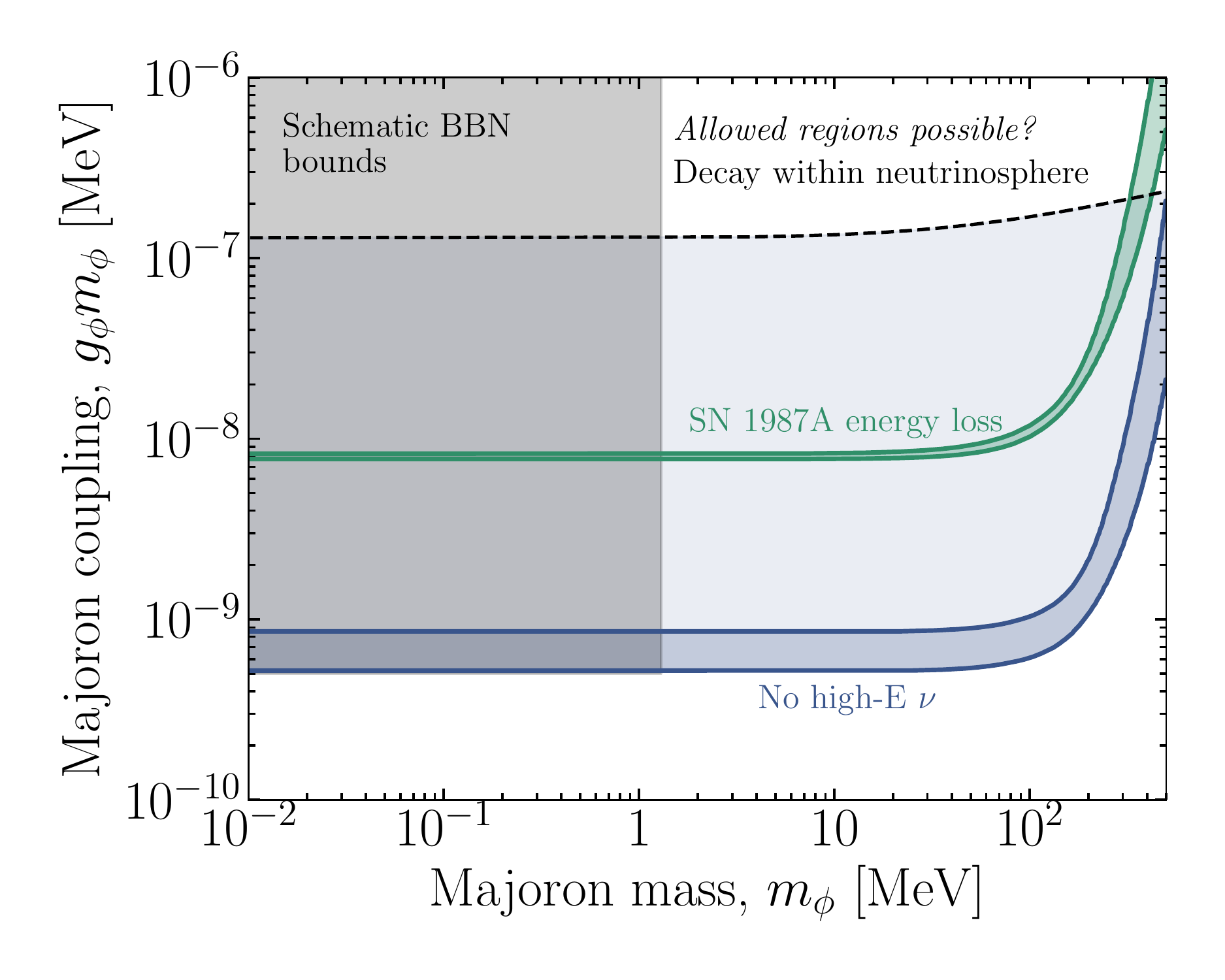}
    \vskip-12pt
    \caption{Constraints on the Majoron coupling in the \hbox{$m_\phi$--$g_\phi m_\phi$} plane from SN~1987A energy loss (green) and the absence of 100 MeV-range (``high-E'') events (blue). 
    The shaded range brackets the cold (upper curves) vs.\ hot (lower curves) SN models, i.e., the Garching muonic models SFHo-18.8 and LS220-s20.0 \cite{Bollig:2020xdr}. Above the dashed line, Majorons with a reference kinetic energy of 100~MeV decay before leaving the SN core.    The ``ceiling'' of the energy-loss bound is probably outside
this figure, but we are not confident about its exact location. The schematic big bang nucleosynthesis (BBN) bounds are taken from Fig.~1 of Ref.~\cite{Kelly:2020aks}, based on the cosmic radiation density. Somewhat more restrictive limits may follow from the cosmic microwave background (CMB) (see text).}
    \label{fig:constraints}
\end{figure}

In other cases, FIP decays include active neutrinos. In the free-streaming limit, FIPs escape from the inner SN core and so their decays provide 100-MeV-range events, much larger than the usual neutrino burst of few 10~MeV that emerges from the ``neutrino sphere'' at the edge of the SN core. The background of atmospheric muons has yet larger energies and so the new signal would stick out in a future SN neutrino observation. 
This argument was first advanced in Ref.~\cite{Akita:2022etk}, and offers an intriguing future detection opportunity.

Our main point is that, by the same token, SN~1987A already
provides restrictive limits because the legacy data do not sport any events with such intermediate energies. This constraint, which is available today without the need to wait for the next galactic SN, is far more restrictive than the traditional energy-loss argument.

We illustrate our new argument with the simple case of nonstandard or ``secret'' neutrino-neutrino interactions \cite{Heurtier:2016otg, Brune:2018sab, Berryman:2022hds,Akita:2022etk,Chang:2022aas}, mediated by a (pseudo)scalar $\phi$ (mass $m_\phi$) that we call Majoron and take to interact with all flavors with the same strength $g$. We consider $m_\phi\agt100~{\rm eV}$ so that neutrino masses and refractive matter potentials can be ignored. The lepton-number violating production channels $\bar\nu\bar\nu\to\phi$ and $\nu\nu\to\phi$ and corresponding decays yield the constraints previewed in Fig.~\ref{fig:constraints}.

The older Majoron literature \cite{Kolb:1987qy, Aharonov:1988ju, Aharonov:1988ee, Aharonov:1989ik, Fuller:1988ega, Grifols:1988fg, Choi:1987sd, Choi:1989hi, Farzan:2002wx} instead took the low-mass limit where neutrino coalescence $\nu \bar\nu\to\phi$ and decay is enabled by the matter potential and otherwise second-order processes of the type $\nu\phi\to\nu\phi$ or $\nu\bar\nu\to\phi\phi$ dominate. One may consult
Fig.~9 of Ref.~\cite{Berryman:2022hds} for the landscape of constraints, including previous SN~1987A energy-loss limits in our mass range \cite{Heurtier:2016otg,Brune:2018sab}.

{\bf\textit{Majoron decay and production.}}---A universal $\nu$--$\nu$ interaction by Majoron exchange is given by \cite{Farzan:2002wx}
\begin{equation}\label{eq:Lagrangian}
    \mathcal{L}_{\text{int}}=-\frac{g}{2}\,\psi_\nu^T\sigma_2\psi_\nu \phi+\textrm{h.c.},
\end{equation}
where $\psi_\nu$ is a two-component Majorana field and $g$ a real number. In the relativistic limit we refer to the Majorana helicity states as $\nu$ and $\bar\nu$ in the usual sense. 

The decay into pairs of relativistic neutrinos requires equal helicities, implying the lepton-number violating channels $\phi\to\nu\nu$ or $\bar\nu\bar\nu$. Each individual rate is
\begin{equation}
    \Gamma_{\phi\to\nu\nu}=\frac{g^2 m_\phi}{32\pi},
\end{equation}
which includes a symmetry factor $1/2$ for identical final-state particles. (We always use natural units with $\hbar=c=k_{\rm B}=1$.) The total rate requires a factor of 6 for six species~\cite{standardSN}. For a relativistic Majoron, this rate is slower by the Lorentz factor $m_\phi/E_\phi$, implying that the laboratory decay rate depends only on the combination~$g m_\phi$.

The requirement that Majorons with $E_\phi=100$~MeV decay beyond the neutrino-sphere radius of 20~km thus implies $g m_\phi\alt10^{-7}$~MeV, shown as a dashed line in Fig.~\ref{fig:constraints}. On the other hand, the decay neutrinos should not be delayed by more than a few seconds. The requirement $\Gamma^{-1}\alt 1$~s implies $g m_\phi\gtrsim1\times10^{-9}$~MeV for $E_\phi=100$~MeV. The time-of-flight difference is much smaller for relativistic Majorons, so for the constraints shown in Fig.~\ref{fig:constraints} the signals are indeed contemporaneous, although somewhat marginally for $m_\phi$ around 100~MeV.

The neutrino decay spectrum is flat between $E_\pm=\frac{1}{2}\bigl(E_\phi\pm p_\phi\bigr)$ with $p_\phi=(E_\phi^2-m_\phi^2)^{1/2}$. In a neutrino gas of one species $\alpha$, occupation number $f_\alpha(E_\nu)$, the spectral Majoron emission rate from $\nu_\alpha\nu_\alpha$ coalescence then is
\begin{equation}\label{eq:ndot}
    \frac{d\dot{N}_\phi^{(\alpha)}}{dE_\phi}\Big|_{E_\phi}=\frac{g^2 m_\phi^2}{64\pi^3}
    \int_{E_-}^{E_+}\!\!dE_\nu\, f_\alpha (E_\nu) f_\alpha (E_\phi-E_\nu).
\end{equation}
For local thermal equilibrium with temperature $T$ and neutrino chemical potential $\mu_\alpha$, the corresponding Fermi-Dirac distribution is $f_\alpha(E_\nu)=\bigl[e^{(E_\nu-\mu_\alpha)/T}+1\bigr]^{-1}$. The chemical potential for a flavor $\nu_\ell$ enters with opposite sign, depending on $\alpha$ denoting a $\nu$ or $\bar\nu$. Notice that the lepton-number violation caused by the $\phi$ interaction implies $\mu_\nu=0$ in true equilibrium.

All Majorons decay close to the SN equally into all six neutrino species with a flat spectrum. Therefore, the effective single-species spectral neutrino emission rate is
\begin{equation}\label{eq:ndot-nu}
    \frac{d\dot{N}_\alpha}{dE_\nu}\Big|_{E_\nu}=\frac{2}{6}
    \int_{E_{\rm min}}^{\infty}\frac{dE_\phi}{p_\phi}\sum_{\beta=1}^6\frac{d\dot{N}_\phi^{(\beta)}}{dE_\phi}\Big|_{E_\phi}.
\end{equation}
The minimal $E_\phi$ to produce a neutrino of energy $E_\nu$ is $E_{\rm min}=E_\nu+m_\phi^2/4E_\nu$. The first factor of 2 is for two neutrinos per decay, whereas $1/6$ appears because this is the rate into one of six species.

{\bf\textit{One-zone SN model.}}---For a first estimate we use a one-zone model of the collapsed SN core with a chemical potential $\mu_{\nu}=100$~MeV for $\nu_e$ and vanishing for the other flavors, volume $(4\pi/3)R^3$ with $R=10$~km for the emitting region, and duration for substantial deleptonization of $\tau=1$~s~\cite{parameters}\nocite{supplementalmaterial,Haxton:1987kc,Kolbe:2002gk,Scholberg:2012id,Strumia:2003zx,Formaggio:2012cpf,Marteau:1999zp,Langanke:1995he,IAUC4316,1970CoTol..89.....S,Irvine-Michigan-Brookhaven:1983iap,WWVB,VanDerVelde:1989xb,Raffelt:1988gv,Arisaka:1985lki,Kajita:2012zz,Badino:1984ww, Aglietta:1987it,Schaeffer:1987hc,Tamborra:2014aua,Burrows:1986ApJ,Pons:1999ApJ,Li:2020ujl,Pascal:2022MNRAS}. After collapse, the SN core is cold ($T\simeq10$~MeV) and heats up from outside in as the material deleptonizes. Majoron emission is thus from the coalescence of $\nu_e\nu_e$ alone which we take as perfectly degenerate. (In contrast, novel particle emission usually becomes large only after the SN core has heated up at around 1~s after collapse~\cite{Caputo:2021rux}.)

For $m_\phi=0$ the integral in Eq.~\eqref{eq:ndot} is a ``triangle function'' that rises linearly to the value $\mu_\nu$ at $E_\phi=\mu_\nu$ and then decreases linearly to zero at $E_\phi=2\mu_\nu$. The energy-loss rate per unit volume is $Q_\phi=(g m_\phi)^2\mu_\nu^3/64\pi^3$. Comparing $L_\phi=Q_\phi(4\pi/3)R^3$ with $L_\nu\simeq2\times10^{52}~{\rm erg/s}$ as recommended by a simple recipe \cite{Raffelt:2006cw} implies $g m_\phi\alt 4\pi\sqrt{3L_\nu/R^3\mu_\nu^3}=5.5\times10^{-9}~{\rm MeV}$.

Likewise, the effective $\nu_\alpha$ production rate per unit volume is $\dot N_\alpha=(g^2 m_\phi^2/64\pi^3)\,\mu_\nu^2/3$ and therefore the total emitted number is \smash{$N_\alpha=\dot N_\alpha(4\pi/3)R^3\tau$}. The fluence at Earth is $N_\alpha/(4\pi d_{\rm SN}^2)$ where $d_{\rm SN}=49.6$~kpc is the distance to SN~1987A \cite{2019Natur.567..200P}. The largest detector was IMB with a fiducial mass of 6.8~kton \cite{IMB:1988suc} and thus $N_p=4.5\times10^{32}$ fiducial protons. The detection cross section is very roughly $\sigma\simeq \bar\sigma E_\nu^2$ with $\bar\sigma\simeq 10^{-43}~{\rm cm}^2/{\rm MeV}^2$ and $\langle E_\nu^2\rangle=7\mu_\nu^2/18$. The total number of 100-MeV-range events therefore is $N_{e^+}=\sigma N_p N_\alpha/4\pi d_{\rm SN}^2$ and the requirement $N_{e^+}\alt 1$ implies $g m_\phi\alt 72\, (2 d_{\rm SN}^2 \pi^3/7N_p R^3 \mu_\nu^4\bar\sigma \tau)^{1/2}=1\times10^{-9}$~MeV.

{\bf\textit{Numerical SN models.}}---This constraint is much more restrictive than from energy loss, motivating a detailed study. To this end we use the Garching 1D models SFHo-18.8 and LS220-s20.0 that were evolved with the {\sc Prometheus Vertex} code with six-species neutrino transport \cite{JankaWeb}. These muonic models were recently also used for other particle constraints \cite{Bollig:2020xdr, Caputo:2021rux}. With different final neutron-star masses and different equations of state, these models were taken to span the extremes of a cold and a hot case, reaching internal $T$ of around 40 vs.\ 60~MeV. On the other hand, the initial $\mu_{\nu_e}$ profiles are much more similar, in both cases around 150~MeV in the center and a ``lepton core'' reaching up to around 10~km. The lepton number of the outer core layers is released within a few ms after core bounce in the form of the prompt $\nu_e$ burst. More details about these models are provided in the Supplemental Material~\cite{supplementalmaterial}.

SN neutrinos follow a quasi-thermal spectrum that can be represented by a Gamma distribution~\cite{Keil:2002in, Tamborra:2012ac, Vitagliano:2019yzm}. We thus write the time-integrated spectrum in the form
\begin{equation}\label{eq:nufromSN}
    \frac{dN_{\bar{\nu}_e}}{dE_\nu}=\frac{E_{\mathrm{tot}}}{6E_0^2} \frac{(1+\alpha)^{1+\alpha}}{\Gamma(1+\alpha)} \left(\frac{E_\nu}{E_0}\right)^\alpha e^{-(1+\alpha)E_\nu/E_0},
\end{equation}
where $E_{\rm tot}$ is the total SN energy release, $E_0$ the average $\bar\nu_e$ energy, $\alpha$ a parameter that would be 2 for a Maxwell-Boltzmann distribution, and $\Gamma$ the Gamma function, not to be confused with a Gamma distribution. The factor 1/6 represents assumed flavor equipartition. The parameters are chosen such that $E_{\rm tot}$, $E_0=\langle E_\nu\rangle$, and $\langle E_\nu^2\rangle$ agree with the numerical spectrum.

The cold model releases $E_{\rm tot}=1.98\times10^{53}~{\rm erg}$. The exact impact of flavor oscillations on SN neutrinos is not yet fully understood. Averaging over all three $\bar\nu$ flavors, we find $E_0=12.7$~MeV and $\alpha=2.39$. For the hot model, these parameters are $E_{\rm tot}=3.93\times10^{53}~{\rm erg}$, $E_0=14.3$~MeV and $\alpha=2.07$.

{\bf\textit{SN~1987A cooling limit.}}---The local Majoron energy loss follows from Eq.~\eqref{eq:ndot} that we correct for gravitational redshift through the tabulated lapse factors as described in Ref.~\cite{Caputo:2021rux}. In the cold model, we find a Majoron luminosity at 1~s post bounce of $L_\phi(1\,{\rm s})=(g m_{\rm MeV})^2\,6.46\times10^{68}~{\rm erg/s}$, where $m_{\rm MeV}=m_\phi/{\rm MeV}$. According to the traditional SN~1987A cooling argument \cite{Raffelt:2006cw,Caputo:2021rux,scaling} we compare it with $L_\nu(1\,{\rm s})=4.40\times10^{52}~{\rm erg/s}$, leading to $g m_\phi<0.83\times10^{-8}$~MeV shown in Fig.~\ref{fig:constraints}. For larger masses, we include a cutoff for those Majorons that are produced with insufficient energy to escape the gravitational potential as explained in the Supplemental Material of Ref.~\cite{Caputo:2022mah}. The total emission is $E^{\rm tot}_\phi=(g m_{\rm MeV})^2\,1.94\times10^{69}~{\rm erg}$ and nominally $E_\nu^{\rm tot}=E_\phi^{\rm tot}$ for $g m_\phi=0.99\times10^{-8}$~MeV, practically identical to the luminosity comparison at 1~s.

For the hot model we find $L_\phi(1\,{\rm s})=(g m_{\rm MeV})^2\,1.39\times10^{69}~{\rm erg/s}$, to be compared with $L_\nu(1\,{\rm s})=8.29\times10^{52}~{\rm erg/s}$, leading to $g m_\phi<0.77\times10^{-8}$~MeV. Moreover, $E^{\rm tot}_\phi=(g m_{\rm MeV})^2\,4.39\times10^{69}~{\rm erg}$ and $E_\nu^{\rm tot}=E_\phi^{\rm tot}$ for $g m_\phi=0.93\times10^{-8}$~MeV. As seen from these numbers and Fig.~\ref{fig:constraints}, the constraints are very insensitive to the specific SN model and similar to the one-zone estimate.

\begin{figure}
    \centering
    \includegraphics[width=0.95\columnwidth]{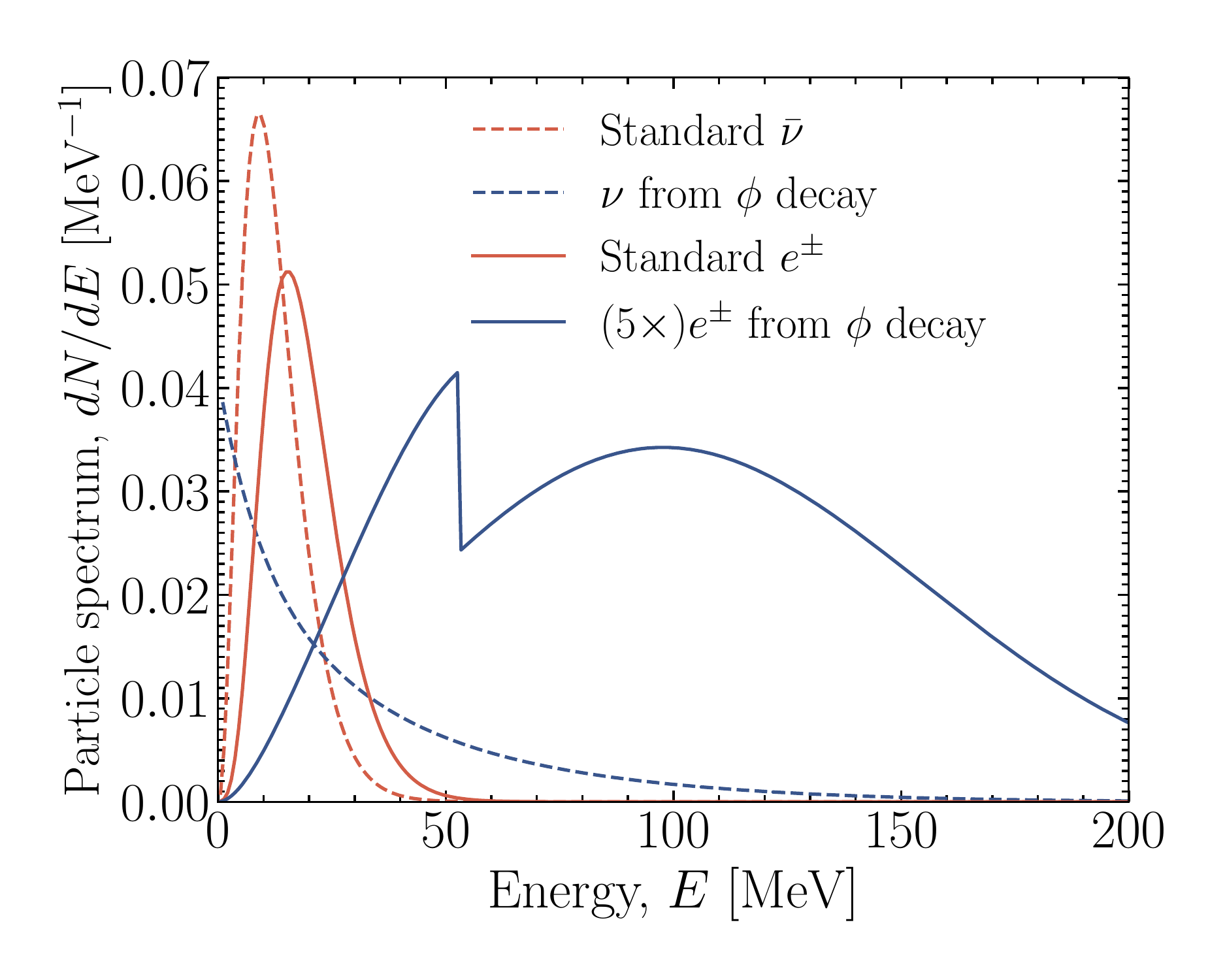}
    \caption{Normalized particle spectra from the time-integrated emission of the cold model SFHo-18.8. ``Standard $\bar\nu$'' is the flavor average of the usual SN $\bar\nu$ and ``Standard $e^\pm$'' the corresponding $e^\pm$ spectrum in the detector (ignoring detection efficiencies), whereas the new contributions are marked ``from $\phi$ decay.'' They include Michel $e^\pm$ (endpoint 53~MeV) from $\mu^\pm$ decays at rest, which themselves emerge from CC interactions of $\nu_\mu$ and $\bar\nu_\mu$ that come from $\phi$ decay.}
    \label{fig:neutrinospectrum}
\end{figure}

{\bf\textit{Neutrino detection.}}---The main SN~1987A neutrino observations came from the water Cherenkov detectors Kamiokande~II (2.14~kton) \cite{Kamiokande-II:1987idp, Hirata:1988ad, Hirata:1991td} and IMB (6.8~kton) \cite{Bionta:1987qt, IMB:1988suc, 1987svoboda}. They observed events with energies up to 40~MeV via inverse beta decay $\bar{\nu}_e + p \to e^+ + n$, whereas elastic scattering on electrons is small (but dominates for solar $\nu_e$ detection). For our 100~MeV-range energies, charged current (CC) reactions on oxygen of the form $\bar{\nu}_{e} + {\rm O} \to e^+ + {\rm X}$ and ${\nu}_{e} + {\rm O} \to e^- + {\rm Y}$ with X and Y excited final-state nuclei, dominate for $E_\nu\agt 70$~MeV. For energies above the muon production threshold ($m_\mu=105.7$~MeV), the corresponding muonic CC processes also happen, especially of course for atmospheric neutrinos at yet larger energies. Muons quickly come to rest by ionization and produce ``Michel $e^{\pm}$'' with a characteristic spectrum ending at 53~MeV, half the muon mass. Below the muon Cherenkov threshold of about 160~MeV, they are termed ``invisible muons.'' (For more details about these processes see the Supplemental Material~\cite{supplementalmaterial}.)

Figure~\ref{fig:neutrinospectrum} shows the spectral fluence (time-integrated flux) for the standard SN neutrinos from the cold model, averaged over $\bar\nu_e$, $\bar\nu_\mu$ and $\bar\nu_\tau$. The energy-integrated fluence is $5.10\times10^9~{\rm cm}^{-2}$ for one species. We also show the corresponding $e^\pm$ spectrum in the detector; the total event number is $5.07$ per kton (for 100\% detection efficiency). Next we show the $\nu$ spectrum from $\phi$ decay which is the same in every species; the total fluence in one species is $(g m_{\rm MeV})^2\,1.90\times10^{25}~{\rm cm}^{-2}$. The $e^\pm$ event number times $(g m_{\rm MeV})^2/{\rm kton}$ is $3.62\times10^{17}$ produced by $\bar\nu_e$ and $\nu_e$ in CC reactions and  $0.37\times10^{17}$ from Michel $e^\pm$ ($E\alt53$~MeV) caused by invisible muons, and a total of $3.99\times10^{17}$.

Above the muon Cherenkov threshold of 160~MeV, and assuming the same detection efficiency as for $e^\pm$, visible $\mu^\pm$ contribute another 11\% to the total events. After each such event, the IMB detector would be blind by trigger dead time, so we should not include the subsequent
Michel events. However, even for $\mu^\pm$ themselves, the Cherenkov threshold behavior and the detection efficiency are not available. Therefore, we do not include visible muons, making our Majoron bounds more conservative by some 5\%.

A single event with 100\% detection efficiency in IMB thus requires $g m_\phi=6.06\times10^{-10}~{\rm MeV}$. For the hot model, the corresponding result is $g m_\phi=3.71\times10^{-10}~{\rm MeV}$, both smaller than the estimate from the one-zone model, where we underestimated the cross section. Once more, the exact SN model is not crucial and we essentially find the limits shown in Fig.~\ref{fig:constraints}.

{\bf\textit{Analysis of SN~1987A data.}}---We now turn to a detailed analysis of the Kamiokande~II and IMB data. We summarize several details in the Supplemental Material~\cite{supplementalmaterial} and here only remark that event information was recorded depending on a hardware trigger. In an off-line analysis, one searched for low-energy few-seconds event clusters. ``Low energy'' was defined in Kamiokande-II as less than 170 photo electrons in the inner detector or $E_e\alt50$~MeV \cite{Kamiokande-II:1987idp, Hirata:1988ad, Hirata:1991td}, whereas IMB used maximally 100 PMTs firing or $E_e\alt75$~MeV \cite{Bionta:1987qt, IMB:1988suc, 1987svoboda}. However, as discussed in Supplemental Material~\cite{supplementalmaterial}, we can conclude that no high-energy events were actually observed even above these thresholds during the SN~1987A burst.

The events from $\phi$ decay overlap with the standard SN signal, so one should perform a maximum likelihood analysis with $g$ and $m_\phi$ as fit parameters. However, the standard SN signal depends on the chosen SN model. For example, our cold (hot) model (using the average $\bar\nu_e$-$\bar\nu_\mu$-$\bar\nu_\tau$ spectrum) would have produced 9.12 (21.3) events in Kamiokande~II with average detected electron energy of 20.1 (22.6) MeV, to be compared with the actually observed 12~events with 14.7 MeV average energy. In IMB they would have produced 3.49 (12.5) events on average with 31.3 (34.4) MeV, to be compared with 8~events with 31.9~MeV average. Neither of these models fits the data well and the Kamiokande~II and IMB data are themselves in tension with each other, although in terms of the $E_{\rm tot}$--$E_0$--$\alpha$ parameters one finds credible overlapping values \cite{Jegerlehner:1996kx, Mirizzi:2005tg}.

We do not have a suite of SN models that would allow us to find the one that best fits the SN~1987A data. Instead we represent the signal in the form of Eq.~\eqref{eq:nufromSN} and use an unbinned likelihood for the energies of the events in each detector, as defined in the Supplemental Material~\cite{supplementalmaterial}. First we verify that the maximum of the likelihood for both experiments is at $g=0$, i.e., neither of them prefers the new signal. Next we marginalize the combined likelihood by maximizing it for each value of $g$ and $m_\phi$ over $E_0$ and $E_\mathrm{tot}$. This guarantees our constraints to be conservative, because for each choice of the Majoron parameters we choose the SN neutrino spectral shape as the one that maximizes the agreement with the data. We then follow the procedure outlined in Ref.~\cite{Cowan:2010js} to set upper bounds on the Majoron coupling for each value of the Majoron mass; more details on our statistical procedure are given in the Supplemental Material~\cite{supplementalmaterial}.  We show the corresponding constraints, dominated by the IMB data, in Fig.~\ref{fig:constraints}.

{\bf\textit{Discussion and outlook.}}---We have considered FIPs that escape from the inner SN core and later decay into active neutrinos. Our main result is that the lack of 100-MeV-range events in the SN~1987A data provides surprisingly restrictive constraints. Specifically, the energy loss by $\nu\nu\to\phi$ Majoron emission must be less than 1\% of the total binding energy, much more restrictive than the usual SN~1987A cooling limit.

Moreover, our new bound depends mainly on emission during the first second and not on the sparse late-time events or the predicted cooling speed that depends, e.g., on PNS convection. Our result is also insensitive to a concern that the SN~1987A neutron star has not yet been found (see however \cite{Cigan:2019shp,Page:2020gsx}) and that the late events could have been caused by black-hole accretion \cite{Bar:2019ifz}. (See however \cite{Bollig:2020xdr} for a rebuttal of this scenario.)

Our limit implies that the impact on SN physics and the explosion mechanism is small. However, our discussion leaves open what happens for much stronger couplings when Majorons do not freely escape. The SN core could deleptonize already during infall, perhaps preventing a successful explosion. On the other hand, a thermal bounce may still occur \cite{Fuller:1988ega, Rampp:2002kn}. If the interactions are yet stronger, neutrinos and Majorons form a viscous fluid that is more strongly coupled to itself than to the nuclear medium. This peculiar case was recently examined \cite{Chang:2022aas}; the SN~1987A signal may exclude a certain range of parameters beyond the upper edge of Fig.~\ref{fig:constraints}. 

For $m_\phi\alt 1$~MeV, the cosmic radiation density measured by BBN provides comparable bounds
(Fig.~1 of Ref.~\cite{Kelly:2020aks}, see also Refs.~\cite{Huang:2017egl, Blinov:2019gcj, Escudero:2019gvw}), and those from the CMB may be more restrictive, but the exact reach in mass and coupling strength was not directly provided. Having different systematic issues, the cosmological and SN~1987A arguments are nicely complementary for $m_\phi\alt1$~MeV, whereas the SN~1987A sensitivity is unique for larger~$m_\phi$.

Our method can be applied to any class of FIPs decaying to neutrinos. Examples include heavy neutral leptons \cite{Fuller:2008erj,Magill:2018jla} and 
gauge bosons arising from new symmetries like $U(1)_{L_\mu{-}L_\tau}$ \cite{Foot:1990mn,He:1991qd}, which can be further constrained relative to
the existing bounds from energy loss~\cite{Escudero:2019gzq,Croon:2020lrf}. Notice also that bosons coupling \textit{exclusively} to neutrinos have different production rates if the coalescence process is lepton-number conserving ($\nu\bar\nu\to\phi$) or violating ($\nu\nu\to\phi$)
because in the PNS core, the neutrino and antineutrino distributions differ.

At present it remains open if there exist allowed Majoron parameters somewhere in the trapping regime, a question left for future study. Couplings below our limit leave open the exciting possibility of a detection in the neutrino signal of a future galactic SN \cite{Akita:2022etk} that would reveal FIP emission from the inner SN core.

{\bf\textit{Note Added.}}---Since our paper had appeared on arXiv, our new argument
was used to constrain the heavy-lepton model of Ref.~\cite{Brdar:2023tmi}.

{\bf\textit{Acknowledgements.}}---We are indebted to M.~Nakahata and T.\ Kajita for sharing unpublished information about the Kamiokande-II legacy data, and likewise J.~Learned, J.~LoSecco, and R.~Svoboda for the analogous information about IMB. We thank H.-T.~Janka and R.~Bollig for providing the SN profiles used for our numerical estimates. GR acknowledges support by the German Research Foundation (DFG) through the Collaborative Research Centre ``Neutrinos and Dark Matter in Astro and Particle Physics (NDM),'' Grant SFB-1258, and under Germany's Excellence Strategy through the Cluster of Excellence ORIGINS EXC-2094-390783311. DFGF is supported by the Villum Fonden under Project No.\ 29388 and the European Union's Horizon 2020 Research and Innovation Program under the Marie Sk{\l}odowska-Curie Grant Agreement No.\ 847523 ``INTERACTIONS.'' EV thanks the Niels Bohr Institute for hospitality, and acknowledges support by the US Department of Energy (DOE) Grant DE-SC0009937, the Rosenfeld Foundation, and the Carlsberg Foundation (CF18-0183).

\bibliographystyle{bibi}
\bibliography{References}

\providecommand{\href}[2]{#2}\begingroup\raggedright\begin{thebibliography}{10}

\bibitem{Raffelt:1996wa}
G.~G. Raffelt, \emph{{Stars as laboratories for fundamental physics}}.
  University of Chicago Press, 1996.

\bibitem{Raffelt:2006cw}
G.~G. Raffelt, \emph{{Astrophysical axion bounds}},
  \href{https://doi.org/10.1007/978-3-540-73518-2_3}{\emph{Lect. Notes Phys.}
  {\bfseries 741} (2008) 51}
  [\href{https://arxiv.org/abs/hep-ph/0611350}{{\ttfamily hep-ph/0611350}}].

\bibitem{DiLuzio:2021ysg}
L.~Di~Luzio, M.~Fedele, M.~Giannotti, F.~Mescia and E.~Nardi, \emph{{Stellar
  evolution confronts axion models}},
  \href{https://doi.org/10.1088/1475-7516/2022/02/035}{\emph{JCAP} {\bfseries
  02} (2022) 035} [\href{https://arxiv.org/abs/2109.10368}{{\ttfamily
  2109.10368}}].

\bibitem{Heurtier:2016otg}
L.~Heurtier and Y.~Zhang, \emph{{Supernova Constraints on Massive
  (Pseudo)Scalar Coupling to Neutrinos}},
  \href{https://doi.org/10.1088/1475-7516/2017/02/042}{\emph{JCAP} {\bfseries
  02} (2017) 042} [\href{https://arxiv.org/abs/1609.05882}{{\ttfamily
  1609.05882}}].

\bibitem{Brune:2018sab}
T.~Brune and H.~P\"as, \emph{{Massive Majorons and constraints on the
  Majoron-neutrino coupling}},
  \href{https://doi.org/10.1103/PhysRevD.99.096005}{\emph{Phys. Rev. D}
  {\bfseries 99} (2019) 096005}
  [\href{https://arxiv.org/abs/1808.08158}{{\ttfamily 1808.08158}}].

\bibitem{Berryman:2022hds}
J.~M. Berryman et~al., \emph{{Neutrino self-interactions: A white paper}},
  \href{https://doi.org/10.1016/j.dark.2023.101267}{\emph{Phys. Dark Univ.}
  {\bfseries 42} (2023) 101267}
  [\href{https://arxiv.org/abs/2203.01955}{{\ttfamily 2203.01955}}]. {(2022
  Snowmass Summer Study)}.

\bibitem{Akita:2022etk}
K.~Akita, S.~H. Im and M.~Masud, \emph{{Probing non-standard neutrino
  interactions with a light boson from next galactic and diffuse supernova
  neutrinos}}, \href{https://doi.org/10.1007/JHEP12(2022)050}{\emph{JHEP}
  {\bfseries 12} (2022) 050}
  [\href{https://arxiv.org/abs/2206.06852}{{\ttfamily 2206.06852}}].

\bibitem{Chang:2022aas}
P.-W. Chang, I.~Esteban, J.~F. Beacom, T.~A. Thompson and C.~M. Hirata,
  \emph{{Towards Powerful Probes of Neutrino Self-Interactions in Supernovae}},
   \href{https://arxiv.org/abs/2206.12426}{{\ttfamily 2206.12426}}.

\bibitem{Kamiokande-II:1987idp}
{\scshape Kamiokande-II} Collaboration, K.~Hirata et~al., \emph{{Observation of
  a Neutrino Burst from the Supernova SN1987A}},
  \href{https://doi.org/10.1103/PhysRevLett.58.1490}{\emph{Phys. Rev. Lett.}
  {\bfseries 58} (1987) 1490}.

\bibitem{Hirata:1988ad}
K.~S. Hirata et~al., \emph{{Observation in the Kamiokande-II detector of the
  neutrino burst from supernova SN1987A}},
  \href{https://doi.org/10.1103/PhysRevD.38.448}{\emph{Phys. Rev. D} {\bfseries
  38} (1988) 448}.

\bibitem{Hirata:1991td}
K.~S. Hirata, \emph{{Search for supernova neutrinos at Kamiokande-II}}, Ph.D.
  thesis, Tokyo University, 1991.
\newblock
  \href{https://lib-extopc.kek.jp/preprints/PDF/1991/9105/9105084.pdf}{Link to
  KEK repository}.

\bibitem{Koshiba:1992yb}
M.~Koshiba, \emph{{Observational neutrino astrophysics}},
  \href{https://doi.org/10.1016/0370-1573(92)90083-C}{\emph{Phys. Rept.}
  {\bfseries 220} (1992) 229}.

\bibitem{Oyama:2021oqp}
Y.~Oyama, \emph{{Re-examination of the Time Structure of the SN1987A Neutrino
  Burst Data in Kamiokande-II}},
  \href{https://doi.org/10.3847/1538-4357/ac374b}{\emph{Astrophys. J.}
  {\bfseries 922} (2021) 223}
  [\href{https://arxiv.org/abs/2108.01783}{{\ttfamily 2108.01783}}].

\bibitem{Bionta:1987qt}
R.~M. Bionta et~al., \emph{{Observation of a Neutrino Burst in Coincidence with
  Supernova SN 1987A in the Large Magellanic Cloud}},
  \href{https://doi.org/10.1103/PhysRevLett.58.1494}{\emph{Phys. Rev. Lett.}
  {\bfseries 58} (1987) 1494}.

\bibitem{IMB:1988suc}
{\scshape IMB} Collaboration, C.~B. Bratton et~al., \emph{{Angular distribution
  of events from SN1987A}},
  \href{https://doi.org/10.1103/PhysRevD.37.3361}{\emph{Phys. Rev. D}
  {\bfseries 37} (1988) 3361}.

\bibitem{1987svoboda}
{\scshape IMB} Collaboration, R.~Svoboda et~al.,
  \emph{\href{https://ui.adsabs.harvard.edu/abs/1987ESOC...26..229S}{Neutrinos
  from Supernova 1987A in the IMB Detector}},
\newblock in: ESO Workshop on the SN 1987A, Garching, July 6--8, 1987.

\bibitem{Alekseev:1987ej}
E.~N. Alekseev, L.~N. Alekseeva, V.~I. Volchenko and I.~V. Krivosheina,
  \emph{{Possible Detection of a Neutrino Signal on 23 February 1987 at the
  Baksan Underground Scintillation Telescope of the Institute of Nuclear
  Research}}, {\emph{JETP Lett.} {\bfseries 45} (1987) 589}.
  \href{http://jetpletters.ru/ps/1245/article_18825.pdf}{http://jetpletters.ru/ps/1245/article{\_}18825.pdf}.

\bibitem{Alekseev:1988gp}
E.~N. Alekseev, L.~N. Alekseeva, I.~V. Krivosheina and V.~I. Volchenko,
  \emph{{Detection of the neutrino signal from {SN~1987A} in the {LMC} using
  the INR Baksan Underground Scintillation Telescope}},
  \href{https://doi.org/10.1016/0370-2693(88)91651-6}{\emph{Phys. Lett. B}
  {\bfseries 205} (1988) 209}.

\bibitem{Caputo:2022rca}
A.~Caputo, G.~Raffelt and E.~Vitagliano, \emph{{Radiative transfer in stars by
  feebly interacting bosons}},
  \href{https://doi.org/10.1088/1475-7516/2022/08/045}{\emph{JCAP} {\bfseries
  08} (2022) 045} [\href{https://arxiv.org/abs/2204.11862}{{\ttfamily
  2204.11862}}].

\bibitem{Caputo:2022mah}
A.~Caputo, H.-T. Janka, G.~Raffelt and E.~Vitagliano, \emph{{Low-Energy
  Supernovae Severely Constrain Radiative Particle Decays}},
  \href{https://doi.org/10.1103/PhysRevLett.128.221103}{\emph{Phys. Rev. Lett.}
  {\bfseries 128} (2022) 221103}
  [\href{https://arxiv.org/abs/2201.09890}{{\ttfamily 2201.09890}}].

\bibitem{Giannotti:2010ty}
M.~Giannotti, L.~D. Duffy and R.~Nita, \emph{{New constraints for heavy
  axion-like particles from supernovae}},
  \href{https://doi.org/10.1088/1475-7516/2011/01/015}{\emph{JCAP} {\bfseries
  01} (2011) 015} [\href{https://arxiv.org/abs/1009.5714}{{\ttfamily
  1009.5714}}].

\bibitem{Oberauer:1993yr}
L.~Oberauer, C.~Hagner, G.~Raffelt and E.~Rieger, \emph{{Supernova bounds on
  neutrino radiative decays}},
  \href{https://doi.org/10.1016/0927-6505(93)90004-W}{\emph{Astropart. Phys.}
  {\bfseries 1} (1993) 377}.

\bibitem{Jaeckel:2017tud}
J.~Jaeckel, P.~C. Malta and J.~Redondo, \emph{{Decay photons from the axionlike
  particles burst of type II supernovae}},
  \href{https://doi.org/10.1103/PhysRevD.98.055032}{\emph{Phys. Rev. D}
  {\bfseries 98} (2018) 055032}
  [\href{https://arxiv.org/abs/1702.02964}{{\ttfamily 1702.02964}}].

\bibitem{Caputo:2021rux}
A.~Caputo, G.~Raffelt and E.~Vitagliano, \emph{{Muonic boson limits: Supernova
  redux}}, \href{https://doi.org/10.1103/PhysRevD.105.035022}{\emph{Phys. Rev.
  D} {\bfseries 105} (2022) 035022}
  [\href{https://arxiv.org/abs/2109.03244}{{\ttfamily 2109.03244}}].

\bibitem{Calore:2020tjw}
F.~Calore, P.~Carenza, M.~Giannotti, J.~Jaeckel and A.~Mirizzi, \emph{{Bounds
  on axionlike particles from the diffuse supernova flux}},
  \href{https://doi.org/10.1103/PhysRevD.102.123005}{\emph{Phys. Rev. D}
  {\bfseries 102} (2020) 123005}
  [\href{https://arxiv.org/abs/2008.11741}{{\ttfamily 2008.11741}}].

\bibitem{Ferreira:2022xlw}
R.~Z. Ferreira, M.~C.~D. Marsh and E.~M\"uller, \emph{{Strong supernovae bounds
  on ALPs from quantum loops}},
  \href{https://doi.org/10.1088/1475-7516/2022/11/057}{\emph{JCAP} {\bfseries
  11} (2022) 057} [\href{https://arxiv.org/abs/2205.07896}{{\ttfamily
  2205.07896}}].

\bibitem{Diamond:2021ekg}
M.~D. Diamond and G.~Marques-Tavares, \emph{{\ensuremath{\gamma}-Ray Flashes
  from Dark Photons in Neutron Star Mergers}},
  \href{https://doi.org/10.1103/PhysRevLett.128.211101}{\emph{Phys. Rev. Lett.}
  {\bfseries 128} (2022) 211101}
  [\href{https://arxiv.org/abs/2106.03879}{{\ttfamily 2106.03879}}].

\bibitem{Caputo:2021kcv}
A.~Caputo, P.~Carenza, G.~Lucente, E.~Vitagliano, M.~Giannotti, K.~Kotake,
  T.~Kuroda and A.~Mirizzi, \emph{{Axionlike Particles from Hypernovae}},
  \href{https://doi.org/10.1103/PhysRevLett.127.181102}{\emph{Phys. Rev. Lett.}
  {\bfseries 127} (2021) 181102}
  [\href{https://arxiv.org/abs/2104.05727}{{\ttfamily 2104.05727}}].

\bibitem{Bollig:2020xdr}
R.~Bollig, W.~DeRocco, P.~W. Graham and H.-T. Janka, \emph{{Muons in
  Supernovae: Implications for the Axion-Muon Coupling}},
  \href{https://doi.org/10.1103/PhysRevLett.125.051104}{\emph{Phys. Rev. Lett.}
  {\bfseries 125} (2020) 051104}
  [\href{https://arxiv.org/abs/2005.07141}{{\ttfamily 2005.07141}}]. [Erratum:
  \href{https://doi.org/10.1103/PhysRevLett.126.189901}{\textit{Phys.Rev.Lett.}
  \textbf{126}, 189901 (2021)}].

\bibitem{Kelly:2020aks}
K.~J. Kelly, M.~Sen and Y.~Zhang, \emph{{Intimate Relationship between Sterile
  Neutrino Dark Matter and \ensuremath{\Delta}Neff}},
  \href{https://doi.org/10.1103/PhysRevLett.127.041101}{\emph{Phys. Rev. Lett.}
  {\bfseries 127} (2021) 041101}
  [\href{https://arxiv.org/abs/2011.02487}{{\ttfamily 2011.02487}}].

\bibitem{Kolb:1987qy}
E.~W. Kolb and M.~S. Turner, \emph{{Supernova 1987A and the secret interactions
  of neutrinos}}, \href{https://doi.org/10.1103/PhysRevD.36.2895}{\emph{Phys.
  Rev. D} {\bfseries 36} (1987) 2895}.

\bibitem{Aharonov:1988ju}
Y.~Aharonov, F.~T. Avignone and S.~Nussinov, \emph{{Neutronization neutrino
  pulses from supernovae and the triplet majoron model}},
  \href{https://doi.org/10.1016/0370-2693(88)91121-5}{\emph{Phys. Lett. B}
  {\bfseries 200} (1988) 122}.

\bibitem{Aharonov:1988ee}
Y.~Aharonov, F.~T. Avignone and S.~Nussinov, \emph{{Implications of the
  triplet-Majoron model for the supernova {SN1987A}}},
  \href{https://doi.org/10.1103/PhysRevD.37.1360}{\emph{Phys. Rev. D}
  {\bfseries 37} (1988) 1360}.

\bibitem{Aharonov:1989ik}
Y.~Aharonov, F.~T. Avignone and S.~Nussinov, \emph{{Comment on ``Constraints on
  the Majoron interactions from the supernova {SN1987A}''}},
  \href{https://doi.org/10.1103/PhysRevD.39.985}{\emph{Phys. Rev. D} {\bfseries
  39} (1989) 985}.

\bibitem{Fuller:1988ega}
G.~M. Fuller, R.~Mayle and J.~R. Wilson, \emph{{The Majoron model and stellar
  collapse}}, \href{https://doi.org/10.1086/166695}{\emph{Astrophys. J.}
  {\bfseries 332} (1988) 826}.

\bibitem{Grifols:1988fg}
J.~A. Grifols, E.~Masso and S.~Peris, \emph{{Majoron couplings to neutrinos and
  {SN1987A}}}, \href{https://doi.org/10.1016/0370-2693(88)91366-4}{\emph{Phys.
  Lett. B} {\bfseries 215} (1988) 593}.

\bibitem{Choi:1987sd}
K.~Choi, C.~W. Kim, J.~Kim and W.~P. Lam, \emph{{Constraints on the Majoron
  interactions from the supernova {SN1987A}}},
  \href{https://doi.org/10.1103/PhysRevD.37.3225}{\emph{Phys. Rev. D}
  {\bfseries 37} (1988) 3225}.

\bibitem{Choi:1989hi}
K.~Choi and A.~Santamaria, \emph{{Majorons and supernova cooling}},
  \href{https://doi.org/10.1103/PhysRevD.42.293}{\emph{Phys. Rev. D} {\bfseries
  42} (1990) 293}.

\bibitem{Farzan:2002wx}
Y.~Farzan, \emph{{Bounds on the coupling of the Majoron to light neutrinos from
  supernova cooling}},
  \href{https://doi.org/10.1103/PhysRevD.67.073015}{\emph{Phys. Rev. D}
  {\bfseries 67} (2003) 073015}
  [\href{https://arxiv.org/abs/hep-ph/0211375}{{\ttfamily hep-ph/0211375}}].

\bibitem{standardSN}
As in standard SN theory, ``species'' $\alpha$ denotes a state that could be
  $\nu$ or $\bar\nu$, whereas ``flavor'' denotes any of $\ell=e$, $\mu$, or
  $\tau$.

\bibitem{parameters}
These parameters are roughly calibrated by our cold numerical model. For the
  SN~1987A energy loss argument~\cite{Heurtier:2016otg} or future signal
  predictions for 100-MeV-range events~\cite{Akita:2022etk}, instead the values
  $\mu_\nu=200~{\rm MeV}$, $R=10~{\rm km}$, and $\tau=10$~s were used, leading
  to overly restrictive limits and overly ambitious signal predictions (see
  Sec.~E of the Supplemental Material~\cite{supplementalmaterial}).

\bibitem{supplementalmaterial}
See Supplemental Material for further details about the detection cross
  sections for SN neutrinos in a water Cherenkov detector used in our analysis,
  the historical SN 1987A observations, our statistical analysis, and the
  Garching SN models. It includes
  Refs.~\cite{Haxton:1987kc,Kolbe:2002gk,Scholberg:2012id,Strumia:2003zx,Formaggio:2012cpf,Marteau:1999zp,Langanke:1995he,IAUC4316,1970CoTol..89.....S,Irvine-Michigan-Brookhaven:1983iap,WWVB,VanDerVelde:1989xb,Raffelt:1988gv,Arisaka:1985lki,Kajita:2012zz,Badino:1984ww,
  Aglietta:1987it,Schaeffer:1987hc,Tamborra:2014aua,Burrows:1986ApJ,Pons:1999ApJ,Li:2020ujl,Pascal:2022MNRAS}.

\bibitem{Haxton:1987kc}
W.~C. Haxton, \emph{{The nuclear response of water Cherenkov detectors to
  supernova and solar neutrinos}},
  \href{https://doi.org/10.1103/PhysRevD.36.2283}{\emph{Phys. Rev. D}
  {\bfseries 36} (1987) 2283}.

\bibitem{Kolbe:2002gk}
E.~Kolbe, K.~Langanke and P.~Vogel, \emph{{Estimates of weak and
  electromagnetic nuclear decay signatures for neutrino reactions in
  Super-Kamiokande}},
  \href{https://doi.org/10.1103/PhysRevD.66.013007}{\emph{Phys. Rev. D}
  {\bfseries 66} (2002) 013007}.

\bibitem{Scholberg:2012id}
K.~Scholberg, \emph{{Supernova Neutrino Detection}},
  \href{https://doi.org/10.1146/annurev-nucl-102711-095006}{\emph{Ann. Rev.
  Nucl. Part. Sci.} {\bfseries 62} (2012) 81}
  [\href{https://arxiv.org/abs/1205.6003}{{\ttfamily 1205.6003}}].

\bibitem{Strumia:2003zx}
A.~Strumia and F.~Vissani, \emph{{Precise quasielastic neutrino/nucleon
  cross-section}},
  \href{https://doi.org/10.1016/S0370-2693(03)00616-6}{\emph{Phys. Lett. B}
  {\bfseries 564} (2003) 42}
  [\href{https://arxiv.org/abs/astro-ph/0302055}{{\ttfamily
  astro-ph/0302055}}].

\bibitem{Formaggio:2012cpf}
J.~A. Formaggio and G.~P. Zeller, \emph{{From eV to EeV: Neutrino Cross
  Sections Across Energy Scales}},
  \href{https://doi.org/10.1103/RevModPhys.84.1307}{\emph{Rev. Mod. Phys.}
  {\bfseries 84} (2012) 1307}
  [\href{https://arxiv.org/abs/1305.7513}{{\ttfamily 1305.7513}}].

\bibitem{Marteau:1999zp}
J.~Marteau, J.~Delorme and M.~Ericson, \emph{{Neutrino oxygen interactions:
  Role of nuclear physics in the atmospheric neutrino anomaly}},  in
  \emph{{34th Rencontres de Moriond: Electroweak Interactions and Unified
  Theories}}, {\vietnam{Thế Giới} Publishers, Vietnam}, 1999.

\bibitem{Langanke:1995he}
K.~Langanke, P.~Vogel and E.~Kolbe, \emph{{Signal for supernova muon-neutrino
  and tau-neutrino neutrinos in water Cherenkov detectors}},
  \href{https://doi.org/10.1103/PhysRevLett.76.2629}{\emph{Phys. Rev. Lett.}
  {\bfseries 76} (1996) 2629}
  [\href{https://arxiv.org/abs/nucl-th/9511032}{{\ttfamily nucl-th/9511032}}].

\bibitem{IAUC4316}
\emph{{IAUC 4316}},  \url{http://www.cbat.eps.harvard.edu/iauc/04300/04316.html
  }.

\bibitem{1970CoTol..89.....S}
N.~{Sanduleak}, \emph{{A deep objective-prism survey for Large Magellanic Cloud
  members}}, {\emph{Contributions from the Cerro Tololo Inter-American
  Observatory} {\bfseries 89} (1970) }.

\bibitem{Irvine-Michigan-Brookhaven:1983iap}
{\scshape IMB} Collaboration, R.~M. Bionta et~al., \emph{{A Search for Proton
  Decay Into $e^+ \pi^0$}},
  \href{https://doi.org/10.1103/PhysRevLett.51.27}{\emph{Phys. Rev. Lett.}
  {\bfseries 51} (1983) 27}. [Erratum:
  \href{https://doi.org/10.1103/PhysRevLett.51.522.4}{\textit{Phys. Rev. Lett.}
  \textbf{51}, 522 (1983)}].

\bibitem{WWVB}
\url{https://www.nist.gov/pml/time-and-frequency-division/time-distribution/radio-station-wwvb}.

\bibitem{VanDerVelde:1989xb}
J.~C. van~der Velde, \emph{{Possible evidence for a new particle from SN
  1987A}}, \href{https://doi.org/10.1103/PhysRevD.39.1492}{\emph{Phys. Rev. D}
  {\bfseries 39} (1989) 1492}.

\bibitem{Raffelt:1988gv}
G.~Raffelt, \emph{{Horizontal branch stars and the neutrino signal from SN
  1987A}}, \href{https://doi.org/10.1103/PhysRevD.38.3811}{\emph{Phys. Rev. D}
  {\bfseries 38} (1988) 3811}.

\bibitem{Arisaka:1985lki}
K.~Arisaka et~al., \emph{{Search for Nucleon Decay Into Charged Lepton +
  Mesons}}, \href{https://doi.org/10.1143/JPSJ.54.3213}{\emph{J. Phys. Soc.
  Jap.} {\bfseries 54} (1985) 3213}.

\bibitem{Kajita:2012zz}
T.~Kajita, M.~Koshiba and A.~Suzuki, \emph{{On the origin of the Kamiokande
  experiment and neutrino astrophysics}},
  \href{https://doi.org/10.1140/epjh/e2012-30007-y}{\emph{Eur. Phys. J. H}
  {\bfseries 37} (2012) 33}.

\bibitem{Badino:1984ww}
G.~Badino et~al., \emph{{The 90 ton liquid scintillator detector in the Mont
  Blanc laboratory}}, \href{https://doi.org/10.1007/BF02573783}{\emph{Nuovo
  Cim. C} {\bfseries 7} (1984) 573}.

\bibitem{Aglietta:1987it}
M.~Aglietta et~al., \emph{{On the event observed in the Mont Blanc Underground
  Neutrino observatory during the occurrence of Supernova 1987a}},
  \href{https://doi.org/10.1209/0295-5075/3/12/011}{\emph{EPL} {\bfseries 3}
  (1987) 1315}.

\bibitem{Schaeffer:1987hc}
R.~Schaeffer, Y.~Declais and S.~Jullian, \emph{{The Neutrino Emission of
  {SN1987A}}}, \href{https://doi.org/10.1038/330142a0}{\emph{Nature} {\bfseries
  330} (1987) 142}.

\bibitem{Tamborra:2014aua}
I.~Tamborra, F.~Hanke, H.-T. Janka, B.~M\"uller, G.~G. Raffelt and A.~Marek,
  \emph{{Self-sustained asymmetry of lepton-number emission: A new phenomenon
  during the supernova shock-accretion phase in three dimensions}},
  \href{https://doi.org/10.1088/0004-637X/792/2/96}{\emph{Astrophys. J.}
  {\bfseries 792} (2014) 96} [\href{https://arxiv.org/abs/1402.5418}{{\ttfamily
  1402.5418}}].

\bibitem{Burrows:1986ApJ}
A.~{Burrows} and J.~M. {Lattimer}, \emph{{The Birth of Neutron Stars}},
  \href{https://doi.org/10.1086/164405}{\emph{Astrophys. J.} {\bfseries 307}
  (1986) 178}.

\bibitem{Pons:1999ApJ}
J.~A. {Pons}, S.~{Reddy}, M.~{Prakash}, J.~M. {Lattimer} and J.~A. {Miralles},
  \emph{{Evolution of Proto-Neutron Stars}},
  \href{https://doi.org/10.1086/306889}{\emph{Astrophys. J.} {\bfseries 513}
  (1999) 780} [\href{https://arxiv.org/abs/astro-ph/9807040}{{\ttfamily
  astro-ph/9807040}}].

\bibitem{Li:2020ujl}
S.~W. Li, L.~F. Roberts and J.~F. Beacom, \emph{{Exciting Prospects for
  Detecting Late-Time Neutrinos from Core-Collapse Supernovae}},
  \href{https://doi.org/10.1103/PhysRevD.103.023016}{\emph{Phys. Rev. D}
  {\bfseries 103} (2021) 023016}
  [\href{https://arxiv.org/abs/2008.04340}{{\ttfamily 2008.04340}}].

\bibitem{Pascal:2022MNRAS}
A.~{Pascal}, J.~{Novak} and M.~{Oertel}, \emph{{Proto-neutron star evolution
  with improved charged-current neutrino-nucleon interactions}},
  \href{https://doi.org/10.1093/mnras/stac016}{\emph{Mon. Not. R. Astron. Soc.}
  {\bfseries 511} (2022) 356}
  [\href{https://arxiv.org/abs/2201.01955}{{\ttfamily 2201.01955}}].

\bibitem{2019Natur.567..200P}
G.~{Pietrzy{\'n}ski} et~al., \emph{{A distance to the Large Magellanic Cloud
  that is precise to one per cent}},
  \href{https://doi.org/10.1038/s41586-019-0999-4}{\emph{Nature} {\bfseries
  567} (2019) 200} [\href{https://arxiv.org/abs/1903.08096}{{\ttfamily
  1903.08096}}].

\bibitem{JankaWeb}
\emph{Garching core-collapse supernova research archive},
  \url{https://wwwmpa.mpa-garching.mpg.de/ccsnarchive/}.

\bibitem{Keil:2002in}
M.~T. Keil, G.~G. Raffelt and H.-T. Janka, \emph{{Monte Carlo study of
  supernova neutrino spectra formation}},
  \href{https://doi.org/10.1086/375130}{\emph{Astrophys. J.} {\bfseries 590}
  (2003) 971} [\href{https://arxiv.org/abs/astro-ph/0208035}{{\ttfamily
  astro-ph/0208035}}].

\bibitem{Tamborra:2012ac}
I.~Tamborra, B.~M{\"u}ller, L.~H{\"u}depohl, H.-T. Janka and G.~Raffelt,
  \emph{{High-resolution supernova neutrino spectra represented by a simple
  fit}}, \href{https://doi.org/10.1103/PhysRevD.86.125031}{\emph{Phys. Rev. D}
  {\bfseries 86} (2012) 125031}
  [\href{https://arxiv.org/abs/1211.3920}{{\ttfamily 1211.3920}}].

\bibitem{Vitagliano:2019yzm}
E.~Vitagliano, I.~Tamborra and G.~Raffelt, \emph{{Grand Unified Neutrino
  Spectrum at Earth: Sources and Spectral Components}},
  \href{https://doi.org/10.1103/RevModPhys.92.045006}{\emph{Rev. Mod. Phys.}
  {\bfseries 92} (2020) 45006}
  [\href{https://arxiv.org/abs/1910.11878}{{\ttfamily 1910.11878}}].

\bibitem{scaling}
Scaling the original axion bounds to the Majoron case with that simple recipe
  ignores that here the particle emission is largest directly after core
  bounce, whereas in the axion or similar cases, the core first has to heat up
  and the emission is largest perhaps around 1~s post bounce. Moreover,
  Majorons remove both energy and lepton number. We suspect that the impact on
  the SN~1987A neutrino signal would be larger than implied by scaling the
  axion case. A detailed analysis would require including Majoron losses in
  self-consistent SN models. In view of the much more restrictive counting-rate
  argument, this exercise is not needed and we can post-process existing
  models.

\bibitem{Jegerlehner:1996kx}
B.~Jegerlehner, F.~Neubig and G.~Raffelt, \emph{{Neutrino oscillations and the
  supernova SN1987A signal}},
  \href{https://doi.org/10.1103/PhysRevD.54.1194}{\emph{Phys. Rev. D}
  {\bfseries 54} (1996) 1194}
  [\href{https://arxiv.org/abs/astro-ph/9601111}{{\ttfamily
  astro-ph/9601111}}].

\bibitem{Mirizzi:2005tg}
A.~Mirizzi and G.~G. Raffelt, \emph{{New analysis of the SN 1987A neutrinos
  with a flexible spectral shape}},
  \href{https://doi.org/10.1103/PhysRevD.72.063001}{\emph{Phys. Rev. D}
  {\bfseries 72} (2005) 063001}
  [\href{https://arxiv.org/abs/astro-ph/0508612}{{\ttfamily
  astro-ph/0508612}}].

\bibitem{Cowan:2010js}
G.~Cowan, K.~Cranmer, E.~Gross and O.~Vitells, \emph{{Asymptotic formulae for
  likelihood-based tests of new physics}},
  \href{https://doi.org/10.1140/epjc/s10052-011-1554-0}{\emph{Eur. Phys. J. C}
  {\bfseries 71} (2011) 1554}
  [\href{https://arxiv.org/abs/1007.1727}{{\ttfamily 1007.1727}}]. [Erratum:
  \href{https://doi.org/10.1140/epjc/s10052-011-1554-0}{\textit{Eur.Phys.J.C}
  \textbf{73}, 2501 (2013)}].

\bibitem{Cigan:2019shp}
P.~Cigan et~al., \emph{{High angular resolution ALMA images of dust and
  molecules in the SN 1987A ejecta}},
  \href{https://doi.org/10.3847/1538-4357/ab4b46}{\emph{Astrophys. J.}
  {\bfseries 886} (2019) 51}
  [\href{https://arxiv.org/abs/1910.02960}{{\ttfamily 1910.02960}}].

\bibitem{Page:2020gsx}
D.~Page, M.~V. Beznogov, I.~Garibay, J.~M. Lattimer, M.~Prakash and H.-T.
  Janka, \emph{{NS 1987A in SN 1987A}},
  \href{https://doi.org/10.3847/1538-4357/ab93c2}{\emph{Astrophys. J.}
  {\bfseries 898} (2020) 125}
  [\href{https://arxiv.org/abs/2004.06078}{{\ttfamily 2004.06078}}].

\bibitem{Bar:2019ifz}
N.~Bar, K.~Blum and G.~D'Amico, \emph{{Is there a supernova bound on axions?}},
  \href{https://doi.org/10.1103/PhysRevD.101.123025}{\emph{Phys. Rev. D}
  {\bfseries 101} (2020) 123025}
  [\href{https://arxiv.org/abs/1907.05020}{{\ttfamily 1907.05020}}].

\bibitem{Rampp:2002kn}
M.~Rampp, R.~Buras, H.-T. Janka and G.~Raffelt, \emph{{Core-collapse supernova
  simulations: variations of the input physics}},
  \href{https://arxiv.org/abs/astro-ph/0203493}{{\ttfamily astro-ph/0203493}}.

\bibitem{Huang:2017egl}
G.-Y. Huang, T.~Ohlsson and S.~Zhou, \emph{{Observational constraints on secret
  neutrino interactions from Big Bang Nucleosynthesis}},
  \href{https://doi.org/10.1103/PhysRevD.97.075009}{\emph{Phys. Rev. D}
  {\bfseries 97} (2018) 075009}
  [\href{https://arxiv.org/abs/1712.04792}{{\ttfamily 1712.04792}}].

\bibitem{Blinov:2019gcj}
N.~Blinov, K.~J. Kelly, G.~Z. Krnjaic and S.~D. McDermott, \emph{{Constraining
  the Self-Interacting Neutrino Interpretation of the Hubble Tension}},
  \href{https://doi.org/10.1103/PhysRevLett.123.191102}{\emph{Phys. Rev. Lett.}
  {\bfseries 123} (2019) 191102}
  [\href{https://arxiv.org/abs/1905.02727}{{\ttfamily 1905.02727}}].

\bibitem{Escudero:2019gvw}
M.~Escudero and S.~J. Witte, \emph{{A CMB search for the neutrino mass
  mechanism and its relation to the Hubble tension}},
  \href{https://doi.org/10.1140/epjc/s10052-020-7854-5}{\emph{Eur. Phys. J. C}
  {\bfseries 80} (2020) 294}
  [\href{https://arxiv.org/abs/1909.04044}{{\ttfamily 1909.04044}}].

\bibitem{Fuller:2008erj}
G.~M. Fuller, A.~Kusenko and K.~Petraki, \emph{{Heavy sterile neutrinos and
  supernova explosions}},
  \href{https://doi.org/10.1016/j.physletb.2008.11.016}{\emph{Phys. Lett. B}
  {\bfseries 670} (2009) 281}
  [\href{https://arxiv.org/abs/0806.4273}{{\ttfamily 0806.4273}}].

\bibitem{Magill:2018jla}
G.~Magill, R.~Plestid, M.~Pospelov and Y.-D. Tsai, \emph{{Dipole Portal to
  Heavy Neutral Leptons}},
  \href{https://doi.org/10.1103/PhysRevD.98.115015}{\emph{Phys. Rev. D}
  {\bfseries 98} (2018) 115015}
  [\href{https://arxiv.org/abs/1803.03262}{{\ttfamily 1803.03262}}].

\bibitem{Foot:1990mn}
R.~Foot, \emph{{New Physics From Electric Charge Quantization?}},
  \href{https://doi.org/10.1142/S0217732391000543}{\emph{Mod. Phys. Lett. A}
  {\bfseries 6} (1991) 527}.

\bibitem{He:1991qd}
X.-G. He, G.~C. Joshi, H.~Lew and R.~R. Volkas, \emph{{Simplest $Z'$ model}},
  \href{https://doi.org/10.1103/PhysRevD.44.2118}{\emph{Phys. Rev. D}
  {\bfseries 44} (1991) 2118}.

\bibitem{Escudero:2019gzq}
M.~Escudero, D.~Hooper, G.~Krnjaic and M.~Pierre, \emph{{Cosmology with A Very
  Light $L_{\mu}-L_{\tau}$ Gauge Boson}},
  \href{https://doi.org/10.1007/JHEP03(2019)071}{\emph{JHEP} {\bfseries 03}
  (2019) 071} [\href{https://arxiv.org/abs/1901.02010}{{\ttfamily
  1901.02010}}].

\bibitem{Croon:2020lrf}
D.~Croon, G.~Elor, R.~K. Leane and S.~D. McDermott, \emph{{Supernova Muons: New
  Constraints on $Z$' Bosons, Axions and ALPs}},
  \href{https://doi.org/10.1007/JHEP01(2021)107}{\emph{JHEP} {\bfseries 01}
  (2021) 107} [\href{https://arxiv.org/abs/2006.13942}{{\ttfamily
  2006.13942}}].

\bibitem{Brdar:2023tmi}
V.~Brdar, A.~de~Gouv\^ea, Y.-Y. Li and P.~A.~N. Machado, \emph{{Neutrino
  magnetic moment portal and supernovae: New constraints and multimessenger
  opportunities}},
  \href{https://doi.org/10.1103/PhysRevD.107.073005}{\emph{Phys. Rev. D}
  {\bfseries 107} (2023) 073005}
  [\href{https://arxiv.org/abs/2302.10965}{{\ttfamily 2302.10965}}].

\end{thebibliography}\endgroup

\onecolumngrid
\appendix

\clearpage

\setcounter{equation}{0}
\setcounter{figure}{0}
\setcounter{table}{0}
\setcounter{page}{1}
\makeatletter
\renewcommand{\theequation}{S\arabic{equation}}
\renewcommand{\thefigure}{S\arabic{figure}}
\renewcommand{\thepage}{S\arabic{page}}

\begin{center}
\textbf{\large Supplemental Material for the Paper\\[0.5ex]
Strong Supernova 1987A Constraints on Bosons Decaying to Neutrinos}
\end{center}

We summarize some details about the detection cross sections for SN neutrinos in a water Cherenkov detector used in our analysis, the historical SN~1987A observations, our statistical analysis, and the Garching SN models.

\bigskip

\twocolumngrid

\section{A.~Detection cross sections}

The primary channel for neutrino detection from SN 1987A was inverse beta decay (IBD) $\bar{\nu}_e + p \to e^+ + n$ on the hydrogen nuclei of the water molecules. Neglecting the recoil of the nucleus, the final positron has an energy $E_e=E_\nu-Q_{\bar{\nu}_e p}$, with $Q_{\bar{\nu}_e p}=1.29$~MeV, and it emits Cherenkov radiation visible in the detector. At typical SN energies, $\bar{\nu}_\mu$ and $\bar{\nu}_\tau$ are kinematically unable to interact via charged current (CC).

Above about 70~MeV, neutrino interactions in a water Cherenkov detector start to be dominated by CC reactions on oxygen of the form $\nu_e+{}^{16}{\rm O}\to e^- + {\rm X}$, where X is a final excited nuclear state dominated by ${}^{16}{\rm F}^*$ \cite{Haxton:1987kc,Kolbe:2002gk,Scholberg:2012id} and a similar reaction for antineutrinos, where the dominant final state is ${}^{16}{\rm N}^*$. The final state $e^\pm$ retains memory of the initial neutrino energy. Specifically we use $E_{e^-}=E_\nu-Q_{\nu_e {\rm O}}$, with $Q_{\nu_e {\rm O}}=15.4$~MeV, and the positron energy is $E_{e^+}=E_{\bar{\nu}}-Q_{\bar{\nu}_e {\rm O}}$ with $Q_{\bar{\nu}_e {\rm O}}=11.4$~MeV.

\begin{figure}
    \centering
    \includegraphics[width=0.48\textwidth]{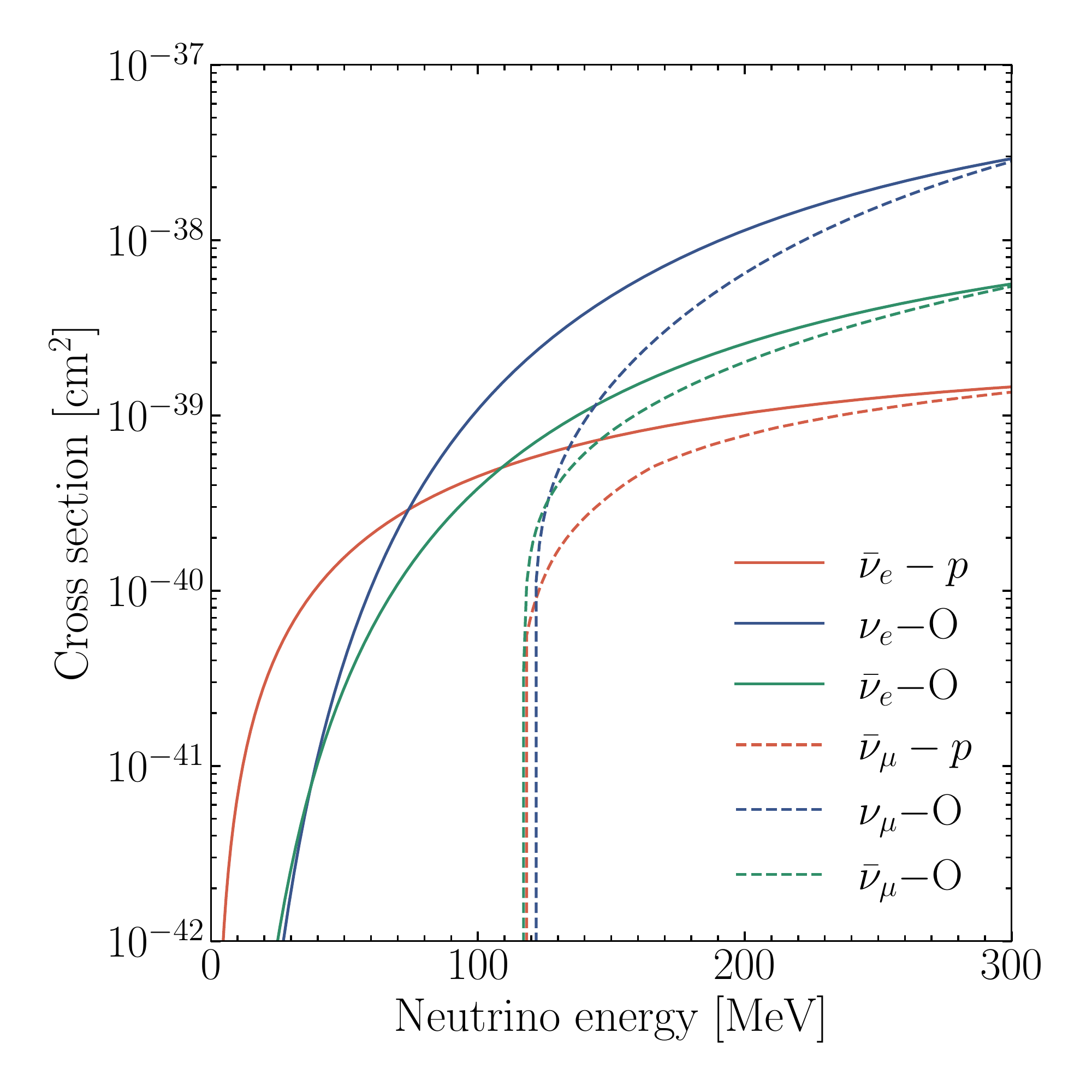}
    \caption{Charged-current neutrino cross sections in a water Cherenkov detector.
    }
    \label{fig:crosssections}
\end{figure}

The cross sections are shown in Fig.~\ref{fig:crosssections}, where the one for IBD is taken from Ref.~\cite{Strumia:2003zx}, the one for $\bar{\nu}_\mu p$ scattering from Ref.~\cite{Formaggio:2012cpf}, the ones for $\bar{\nu}_e O$ and $\nu_e O$ from Ref.~\cite{Kolbe:2002gk}, and the ones for $\bar{\nu}_\mu O$ and $\nu_\mu O$ from Ref.~\cite{Marteau:1999zp}.

In this low-energy range, muon and tau neutrinos can only interact with nucleons via neutral-current interactions. In the interaction, nuclei can be excited and promptly decay to photons, leading to a potentially observable signature~\cite{Langanke:1995he}. For a future Galactic SN, this signature is likely to be observed. However, due to the lower cross sections of the neutral-current scattering, this process played no role for SN~1987A and we will not consider it even for our 100-MeV-range neutrinos.

At energies above the muon production threshold ($m_\mu=105.6$~MeV), the muon-flavored neutrinos from $\phi$ decay also contribute to the analogous CC rates.  Due to large energy losses by ionization, these $\mu^\pm$ are stopped within a short length of the order of $1$~m from their interaction vertex, and they finally decay at rest and produce a visible $e^\pm$. They follow the well-known Michel spectrum, 
\begin{equation}
    \frac{dn_e}{dE_e}=\frac{4}{m_\mu}\left(\frac{2E_e}{m_\mu}\right)^2\left(3-\frac{4E_e}{m_\mu}\right).
\end{equation}
It increases with energy up to a sharp cutoff at $E_e=m_\mu/2=53$~MeV.

Above the muon Cherenkov threshold of 160~MeV, they also produce a direct signal that however we do not include. Otherwise we would need to model the threshold behavior and detection efficiency. Leaving out this signal causes only a small and conservative error in the Majoron bounds (see main text).

\section{B.~SN~1987A Neutrino Observations}

Supernova 1987A, in the Large Magellanic Cloud at a distance of $49.59 \pm 0.09_{\rm stat} \pm 0.54_{\rm syst}$ kpc from Earth~\cite{2019Natur.567..200P}, was discovered independently by Ian Shelton, Oscar Duhalde, and Albert Jones~\cite{IAUC4316} on February 23, 1987, and later targeted by searches in the entire electromagnetic spectrum.  The first evidence for optical brightening was found at 10:38 UT (Universal Time) on plates taken by McNaught.  The first naked-eye visible (in the southern hemisphere) SN since the invention of the telescope, its observation is narrated in detail in a review by Koshiba~\cite{Koshiba:1992yb}. This was also the first SN explosion with a known progenitor star, Sanduleak $-69\, 202$, a blue supergiant catalogued by Nicholas Sanduleak in 1970~\cite{1970CoTol..89.....S}. At the time of the explosion, there were four running experiments that were big enough that they could have detected the gargantuan flux of neutrinos emitted in the collapse of a stellar core. 

The largest one was the Irvine-Michigan-Brookhaven (IMB) water Cherenkov detector, an experiment built to look for proton decay \cite{Irvine-Michigan-Brookhaven:1983iap}, that was located in the Morton-Thiokol salt mine (Fairport, Ohio, USA). It was equipped with 2048 8-inch photomultiplier tubes (PMTs) such that 6,800 tons of water (of a total of 8,000 tons) were within the PMT planes, taken as the fiducial volume for the SN~1987A search \cite{IMB:1988suc}. A failure of a high-voltage power supply shortly before SN~1987A left a contiguous quarter of the PMTs off-line with a geometric effect on the trigger efficiency that was later calibrated. The detector was triggered when at least 20~PMTs fired in 50~ns, corresponding to an energy threshold of 15--25~MeV for showering particles \cite{IMB:1988suc}.
(A trigger of 25~PMTs is mentioned in Ref.~\cite{Bionta:1987qt}). The absolute time of an event was recorded to an uncertainty $\pm 50$~ms thanks to the WWVB clock, a time signal radio station operated by the National Institute of Standards and Technology~\cite{WWVB}.

The first IMB event occurred at 7:35:41.374 Universal Time on 23 February 1987, corresponding to 2:35~am local time on a Monday very early morning. 

At the relatively shallow depth of 1570~m water equivalent, the flux of atmospheric muons caused a  trigger rate of 2.7~Hz. Muons are recognized by tracks entering the detector from the outside and of course coming mostly from above. The detector is dead for 35~ms after each trigger. The SN~1987A signal consisted of 8~events and in addition 15 muons were recorded~\cite{1987svoboda}, a total of 23~triggers, amounting to $23\times35~{\rm ms}=0.8~{\rm s}$ dead time, or 13\% of the SN signal duration of 6~s. In Fig.~\ref{fig:efficiencies} we show the geometrically averaged detection efficiency, including the 0.87 reduction by dead time. 

Atmospheric neutrinos are recognized as contained events and occurred at a rate of around 2/day in the energy range 20--2000~MeV~\cite{Bionta:1987qt}. Our new neutrino signal in the 100~MeV range would look like low-energy atmospheric neutrinos.

The SN~1987A burst was found by looking in the recorded data for low-energy few-second event clusters, where ``low energy'' was defined as fewer than 100~PMTs firing, corresponding roughly to a 75~MeV energy cut. However, other than the 8 SN~1987A events and 15 muons, no other triggers occurred that would have been interpreted as a rare atmospheric neutrino.

The SN~1987A events must be due to IBD with a practically isotropic distribution of final-state $e^+$. However, IMB found a conspicuous directional correlation in the opposite direction of SN~1987A, i.e., the events look ``forward peaked.'' This effect is not explained by the detector's geometrical bias due to the 25\% PMT failure. One idea held that the signal was not caused by neutrinos but instead some new $X^0$ bosons that scatter coherently on oxygen and thus generate the observed angular characteristic \cite{VanDerVelde:1989xb}. However, the required cross section is excluded by stellar cooling bounds from the reverse process \cite{Raffelt:1988gv}. No viable explanation other than a rare statistical fluctuation is available. 

With a fiducial mass of 6.8~kton, IMB would have seen the largest number of 100 MeV-range events. At lower energies it suffered from  a trigger efficiency of only 15\% at 20~MeV, but rising to 80\% at 70~MeV. During the SN~1987A burst, no events besides the 15~background~muons + 8 SN~events = 23 triggers were observed.\footnote{R.~Svoboda, J.~Learned and J.~LoSecco, private communication.} We conclude that there were no unreported events above the low-energy criterion of 75~MeV.

\begin{figure}[ht]
    \centering
    \includegraphics[width=0.48\textwidth]{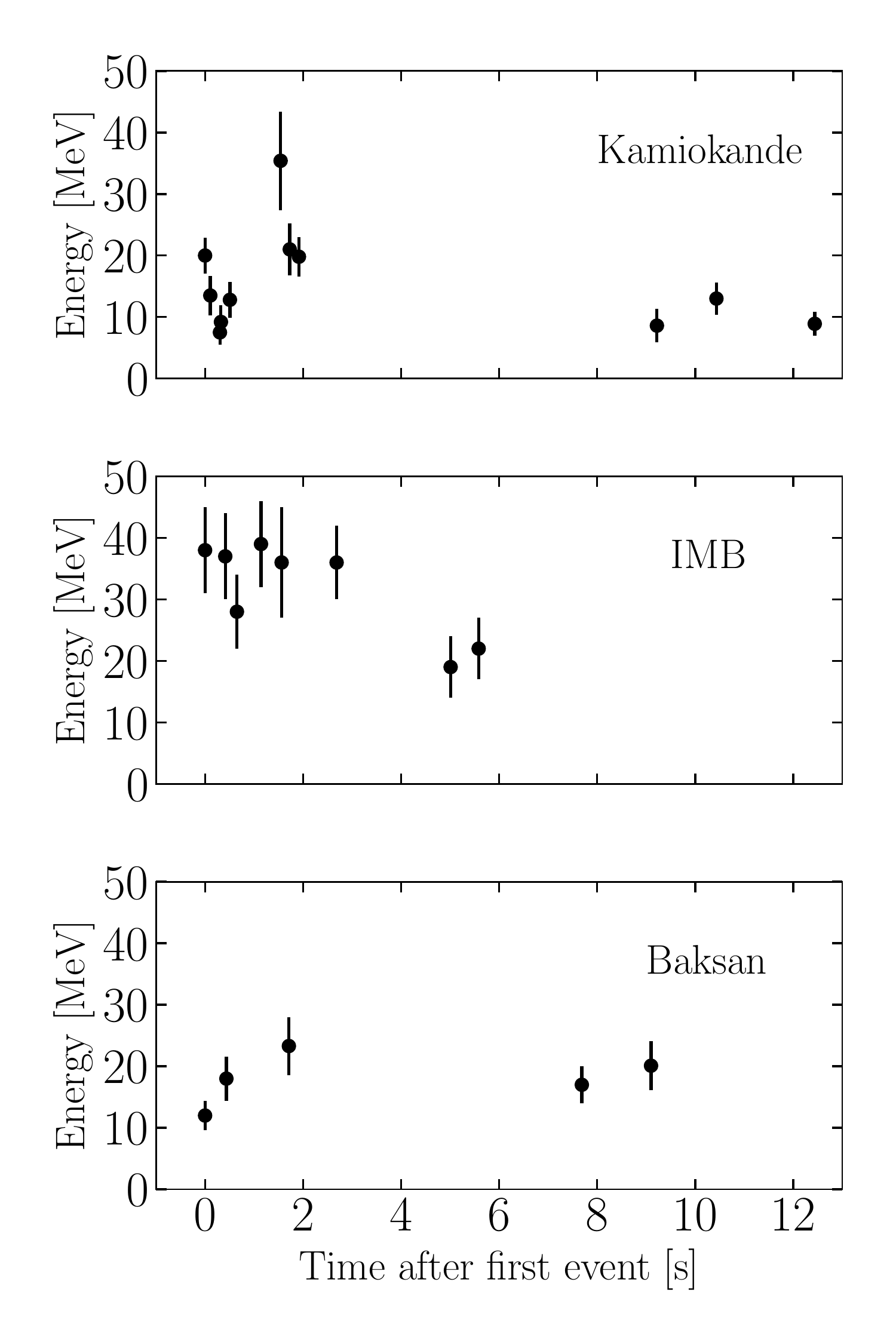}
    \caption{SN~1987A neutrino data collected at Kamiokande-II, IMB, and Baksan. We show the detected positron energy as a function of time after the first event in each detector. Because of clock uncertainties, the exact temporal offset between the observations is not fixed. We do not show events which are attributed to background.}
    \label{fig:SNData}
\end{figure}

\begin{figure}
    \centering
    \includegraphics[width=0.48\textwidth]{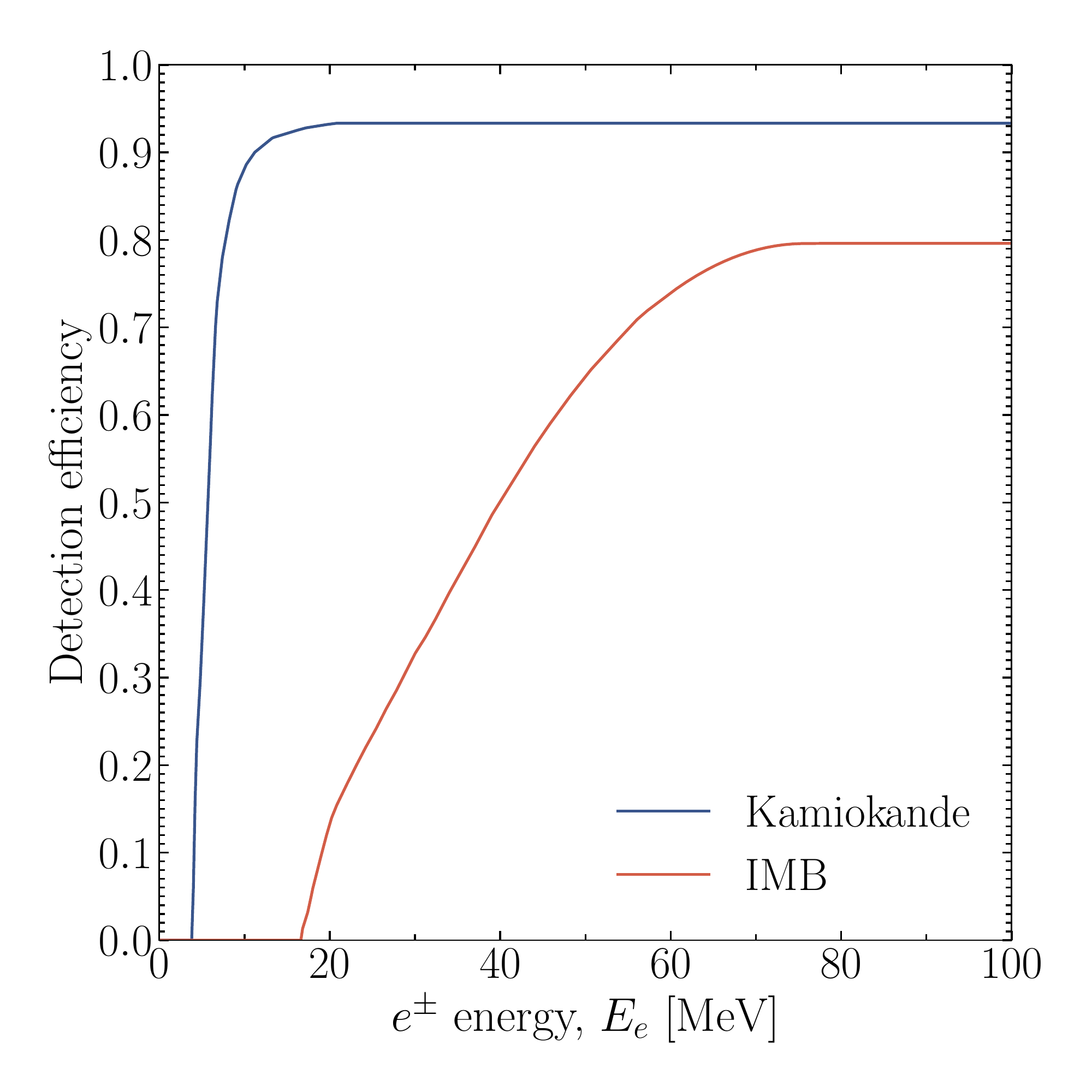}
    \caption{Detection efficiencies for electrons and positrons at Kamiokande and IMB, taken from Ref.~\cite{Jegerlehner:1996kx},  including the dead-time effect in IMB. We continue them at energies higher than $60$~MeV by extrapolation. 
    }
    \label{fig:efficiencies}
\end{figure}

The second largest detector was the Kamiokande II water Cherenkov detector (Mozumi Mine, Kamioka section of Hida, Gifu Prefecture, Japan) with a fiducial mass of 2,140 metric tons for the SN~1987A search, where again the entire volume up to the PMTs was taken \cite{Kamiokande-II:1987idp, Hirata:1988ad}. This detector was built in 1983 to search for nucleon decay (Kamiokande = Kamioka Nucleon Decay Experiment) \cite{Arisaka:1985lki} and later upgraded to Kamiokande II to search for solar $\nu_e$ in the 10~MeV range. The photo cathode coverage was increased and radioactive backgrounds decreased to lower the threshold and solar data were taken since the end of 1986. Despite its smaller mass, the low threshold made Kamiokande II competitive for the SN~1987A discovery (see Fig.~\ref{fig:efficiencies} for the trigger efficiency) although for our 100-MeV-range events, IMB is better suited.

At a greater depth of 2700~m w.e., the atmospheric trigger rate was 0.37~Hz and indeed, 4 muons were found in the 20~s interval preceding the SN~1987A burst and several after the SN burst, but none until just before the 12th event. Atmospheric neutrinos, in the form of fully contained events, show up once every few days. Low-energy radioactive backgrounds triggered with 0.23~Hz. The trigger dead time is less than 50~ns after an event.

To find the SN~1987A burst, the data recorded on a magnetic tape were searched for low-energy event clusters, where here the definition was less than 170 PMTs firing ($E_e\alt 50$~MeV). We show the burst in Fig.~\ref{fig:SNData} as a function of time after the first event. The absolute timing is poorly known, probably to within $\pm 15$~s based on comparing the computer clock with a wrist watch, but a conservative uncertainty of $\pm 1$~min was officially stated. A power outage in the mine on February 26 prevented a recalibration of the computer clock~\cite{Kajita:2012zz}. The signal arrived at 4:35 pm on Monday, 23 February 1987, but this was a substitute holiday. According to working-day schedule, the magnetic tape would have been exchanged at 4:30 pm and the signal might have been missed.

The highest-energy events are also forward peaked, in analogy to IMB, while most of the events are isotropically distributed as expected for IBD. There is a conspicuous gap of 7.3~s between event 9 and 10, filled however with IMB data and probably has nothing to do with SN~1987A. Very recently, one member of the Kamiokande collaboration has speculated that the gap could have been caused by a fault of the magnetic tape drive. He noted that during that gap, there are also no other events (low-energy background or atmospheric muons) and that the probability for such a long gap was very small~\cite{Oyama:2021oqp}.\footnote{However, according to a private communication by M.~Nakahata, this explanation is not viable because the event numbers were continuous across the gap. The event number was generated by the front-end electronics and the trigger system. When the number of hit PMTs within 100~ns was more than a given threshold value, a trigger was generated and the electronics system read out timing and charge information of each individual PMT. The event number was incremented by one whenever a trigger happened. If events had been lost by a tape-write error, there would have to be an event-number gap as well.}

For our analysis, we are mainly interested in the high-energy events that Kamiokande would have seen during the SN~1987A burst. Contained events with 30~MeV $<$ visible energy $<$ 1.33~GeV would have gone into the atmospheric neutrino analysis, but none were found in the period around SN~1987A. For this analysis, the fiducial volume may have been as low as 780~tons (more than 2~m from the wall).\footnote{M.~Nakahata and T.~Kajita, private communication, based on original log books.} We conclude that conservatively no event of our interest was observed in this volume.

The third experiment was the Baksan Scintillator Underground Telescope (BUST) under Mount Andyrchi in the North Caucasus at a depth of 850~m w.e., operated by the Institute of Nuclear Research (Moscow) \cite{Alekseev:1987ej, Alekseev:1988gp}. It started operation in June 1980 and is still running today, with SN~1987A the only SN neutrino burst observed in more than four decades. BUST consists of 3156 segments of $70\times70\times30$~cm. A possible SN~1987A event was selected as one that triggers one and only one segment and with $E_e\alt 50$~MeV. The fiducial inner part has a mass of 130~t that was opened for the SN~1987A analysis to 200~t. Its burst was reported at 7:36:06:571 UT and thus 30~s later than IMB. While the clock synchronization with UT is usually $\pm2$~s, the clock was observed to have shifted forward by 54~s between February 17 and March 11 for unknown reasons. So the observed signal is probably contemporaneous with IMB and Kamiokande~II. Because of its small size, BUST is least useful for us and so we have not investigated how our 100 MeV-range events would have shown up there.

A fourth instrument was the Liquid Scintillation Detector (LSD), located in the gallery of the Mont Blanc tunnel, between Italy and France~\cite{Badino:1984ww, Aglietta:1987it}. It was specifically built to search for a galactic SN burst with a typical assumed distance of 10~kpc. LSD used 72 $100\times 150\times 100 \,\rm cm^3$ liquid scintillator modules, arranged in three horizontal layers for a total mass of 90 tons. Each module was equipped with three PMTs of 15~cm diameter, and the signal was recorded whenever a threefold coincidence occurred within 150~ns.

The LSD collaboration was the first to declare the (possible) discovery of SN neutrinos due to the detection of 5 events, above the 7 MeV threshold, in an interval of 7 seconds, beginning at UT 2:52:36.79 and compatible with the core-collapse standard model at 50~kpc. This signal is almost five hours earlier than the other detectors which observed nothing special at the LSD time and LSD observed nothing special the time of the others. While high multiplicity events can be caused e.g.\ by spallation of oxygen induced by primary muons, no similar event was found during the entire LSD operation which ended with the devastating fire in the Mont Blanc tunnel March 24, 1999. 

The community has settled for the LSD event as being a rare or unexplained fluctuation. A credible physical origin at SN~1987A is astrophysically hard to construct.
Schaeffer, Declais and Jullian computed that, assuming a SN origin for the events seen by LSD, the total energy emitted by SN 1987A would have been $3 \times 10^{54}$ erg, much larger than the value expected by standard core-collapse supernova theory~\cite{Schaeffer:1987hc}.

\section{C.~Statistical Analysis}

We perform our maximum likelihood analysis along the lines of similar previous studies~\cite{Jegerlehner:1996kx, Mirizzi:2005tg}. For the standard SN $\bar\nu_e$ signal we assume a quasi-thermal distribution of the form Eq. (5) described by the three parameters $E_{\rm tot}$, $E_0$ and $\alpha$. We compute the standard $e^+$ signal from the IBD cross section discussed in Sec.~A and for the event spectrum in each detector use the efficiencies discussed earlier, including the IMB dead-time effect of 0.87. 

The SN~1987A are not informative about $\alpha$ \cite{Mirizzi:2005tg}, so we do not try to fit it, but rather use a range of values motivated by numerical SN models. In particular, we use $\alpha=2.39$ (2.07) for the cold (hot) model. The instantaneous neutrino spectra are pinched, i.e., their variance is smaller than that of a Maxwell-Boltzmann spectrum ($\alpha>2$), whereas time-integrated spectra are close to Maxwell Boltzmann. The SN spectra somewhat depend on flavor, but the effect of flavor oscillations is not yet well understood and moreover, because of the LESA effect~\cite{Tamborra:2014aua}, the spectrum depends on the observer direction relative to the 3D structure of the SN explosion.

For each of the two experiments, we thus define an unbinned likelihood
\begin{equation}
    \mathcal{L}(E_0,E_\mathrm{tot})\propto\exp\left[-\!\!\int_{E_{\mathrm{low}}}^{E_{\mathrm{high}}}\! \frac{dN_e}{dE_\mathrm{det}} dE_\mathrm{det}\right]\prod_i \frac{dN_e}{dE_\mathrm{det}}\left(E_i\right), 
\end{equation}
where $E_i$ are the observed energies, and $E_\mathrm{det}$ is the energy reconstructed from the number of firing PMTs, drawn from a Poisson distribution as in Ref.~\cite{Jegerlehner:1996kx}. An unimportant normalization constant has been removed, because we will only deal with likelihood ratios.

In the event rate, we also include the new signal prediction that depends on the parameters $g$ and $m_\phi$; at small masses these appear in the combination $gm_\phi$ and thus collapse to essentially a single parameter. For Kamiokande~II, we reduce the fiducial volume from 2140 tons to 780 tons, as discussed above. We only consider the final-state $e^\pm$ from CC reactions as well as from muon decay, but not the Cherenkov signal caused by muons above the Cherenkov threshold as discussed in the main text. We keep $\alpha$ fixed at the predicted value for the cold and hot SN model. We then marginalize over $E_0$ and $E_{\rm tot}$ as explained in the main text. In this way, we obtain an effective two-dimensiona likelihood
\begin{equation}
    \tilde{\mathcal{L}}(g,m_\phi)=\text{max}_{E_0,E_\mathrm{tot}}(E_0,E_\mathrm{tot},g,m_\phi).
\end{equation}
We now define a test statistic,
\begin{equation}
    \chi^2=2\;\text{max}\left[\log\tilde{\mathcal{L}}(0,m_\phi)-\log\tilde{\mathcal{L}}(g,m_\phi),0\right].
\end{equation}
The asymptotic distribution of this variable under the assumption that Majorons exist is a half-chi-squared distribution~\cite{Cowan:2010js}, which allows us to set a threshold value for 95\% C.L.\ exclusion at $\chi^2=2.7$. With this procedure, we find the limit contours shown in Fig.~1.

\section{D.~Garching Supernova Models}
\label{sec:SN-Models}

In our numerical analysis we use the SN models SFHo-18.8 and LS220-s20.0 from the Garching group that were evolved with the {\sc Prometheus Vertex} code with six-species neutrino transport \cite{JankaWeb} in spherical symmetry. These ``muonic models'' were recently also used for other particle constraints \cite{Bollig:2020xdr, Caputo:2021rux}, where more details are described and radial profiles of various physical quantities are given for specific snapshots of time. PNS convection was taken into account by a mixing-length treatment. Explosions were triggered by hand a few 100~ms after bounce at the Fe/Si or Si/O composition interface of the progenitor star.

Following Ref.~\cite{Bollig:2020xdr}, we note that the SFHo equation of state is fully compatible with all current constraints from nuclear theory and experiment and astrophysics, including pulsar mass measurements and the radius constraints deduced from gravitational-wave and Neutron Star Interior Composition Explorer measurements. For comparison, some of the Garching muonic models also use the traditional LS220 equation of state.

The model SFHo-18.8 \cite{Bollig:2020xdr} uses a progenitor star with mass $18.8\,M_\odot$ that reaches a final neutron-star baryonic mass of $1.351\,M_\odot$ and gravitational mass of $1.241\,M_\odot$, hence a gravitational binding energy of $(1.351-1.241)M_\odot=0.110 M_\odot=1.98\times10^{53}~{\rm erg}$. It is at the lower end of plausible neutron-star masses and released binding energy. It reaches a maximum core temperature near 40~MeV, the coldest of this suite of models. We thus refer to it as our ``cold'' model and it is taken to bracket the lower end of neutron-star mass and core temperature.

The ``hot'' model LS220-s20.0 reaches a maximum core temperature of around 60~MeV. It has a progenitor mass of $20.0\,M_\odot$ and reaches a neutron-star mass of $1.926\,M_\odot$, near the upper end of observed neutron-star masses. Its final gravitational mass is $1.707\,M_\odot$ and thus releases $0.219 M_\odot=3.93\times10^{53}~{\rm erg}$. This model is taken to bracket the upper end of both energy release and internal temperature.

\begin{figure*}
    \centering
    \includegraphics[width=0.32\textwidth]{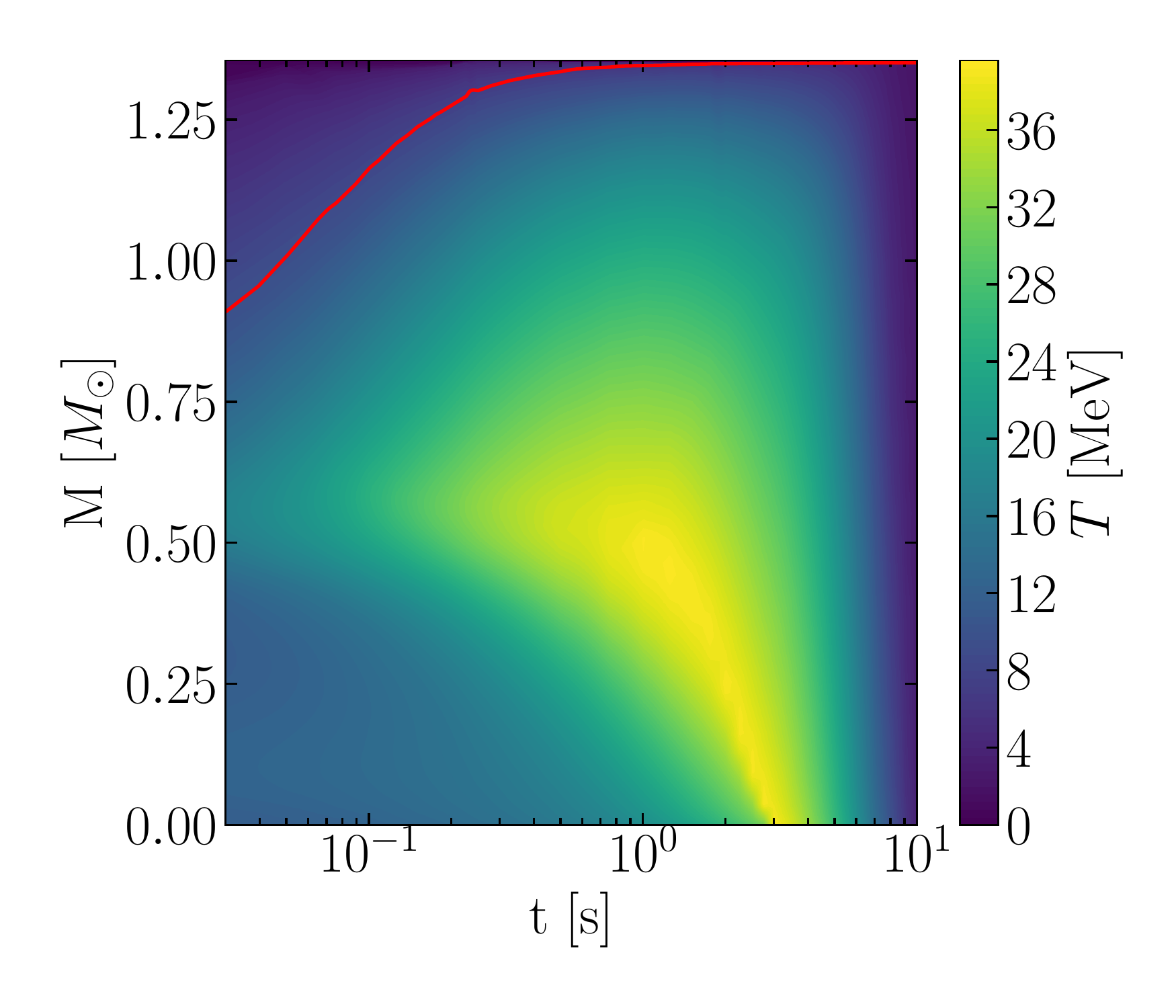}
    \includegraphics[width=0.32\textwidth]{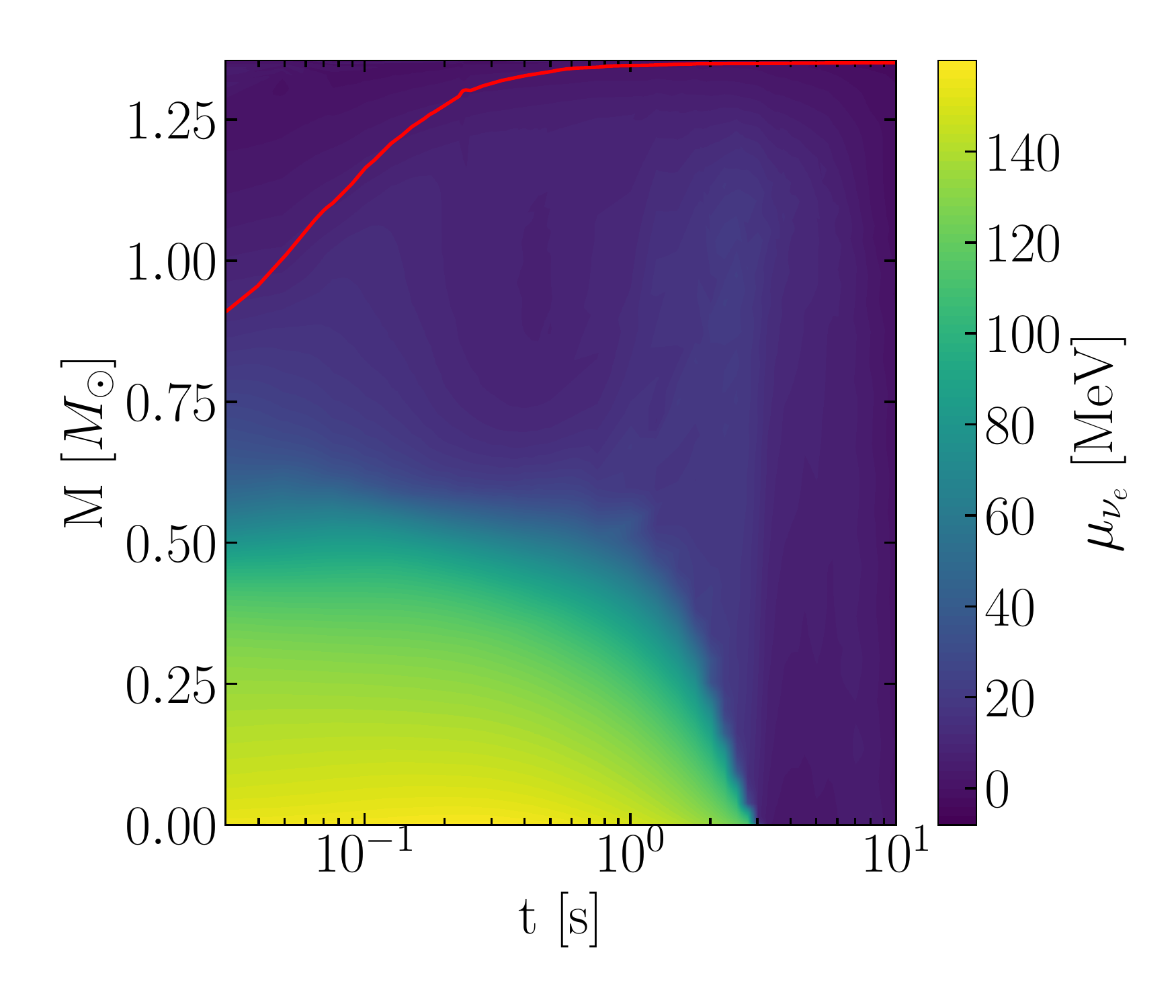}
    \includegraphics[width=0.32\textwidth]{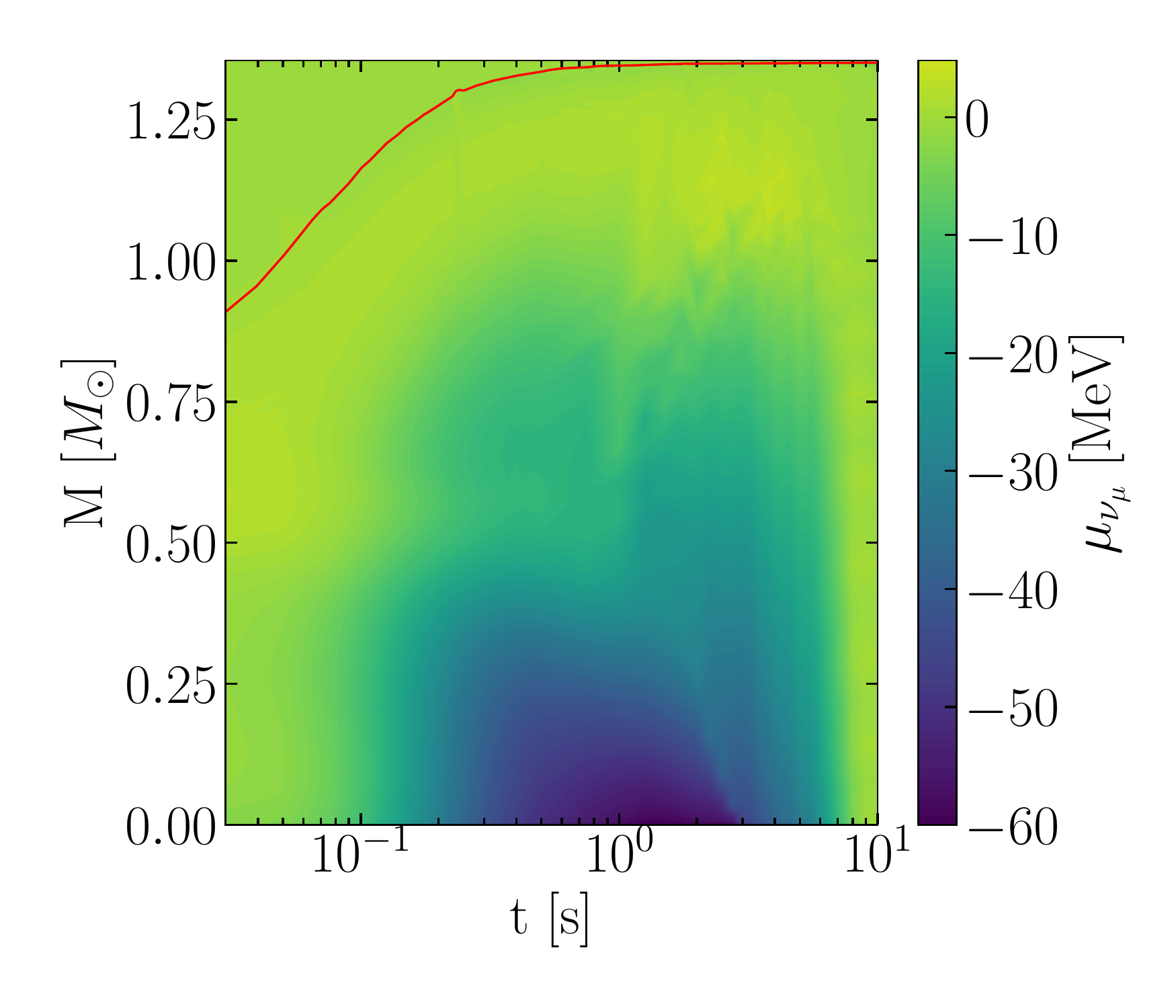}
    \caption{Temperature (left), chemical potential of electron neutrinos (center), and chemical potential of muon neutrinos (right) as a function of post-bounce time and mass coordinate for the Garching ``cold'' model. The red line identifies the density $3\times 10^{12}$~g~cm$^{-3}$ and thus essentially the edge of the PNS. The final neutron-star mass is $1.351\,M_\odot$.}
    \label{fig:coldmodelproperties}
\vskip12pt
    \includegraphics[width=0.32\textwidth,height=0.32\textwidth]{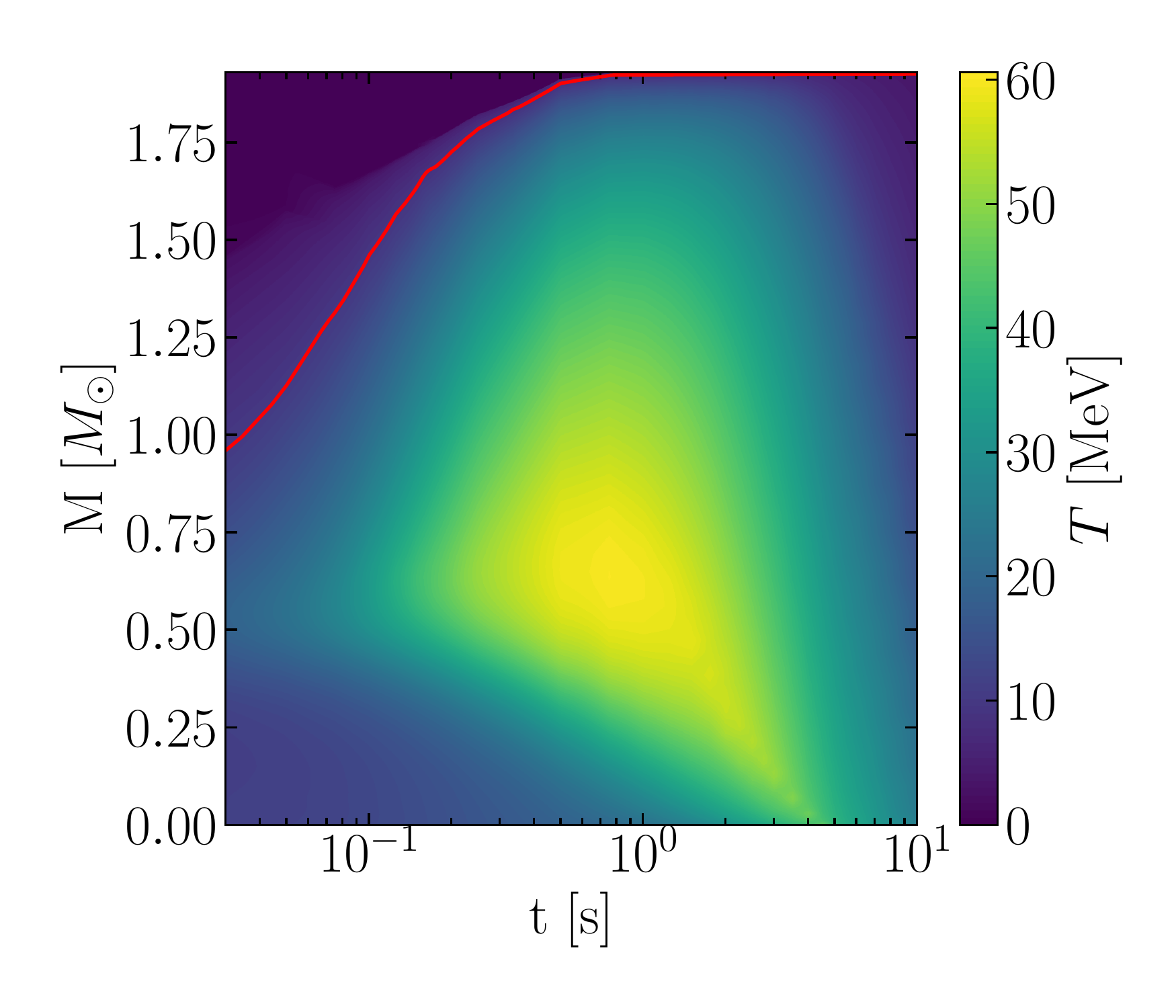}
    \includegraphics[width=0.32\textwidth,height=0.3201\textwidth]{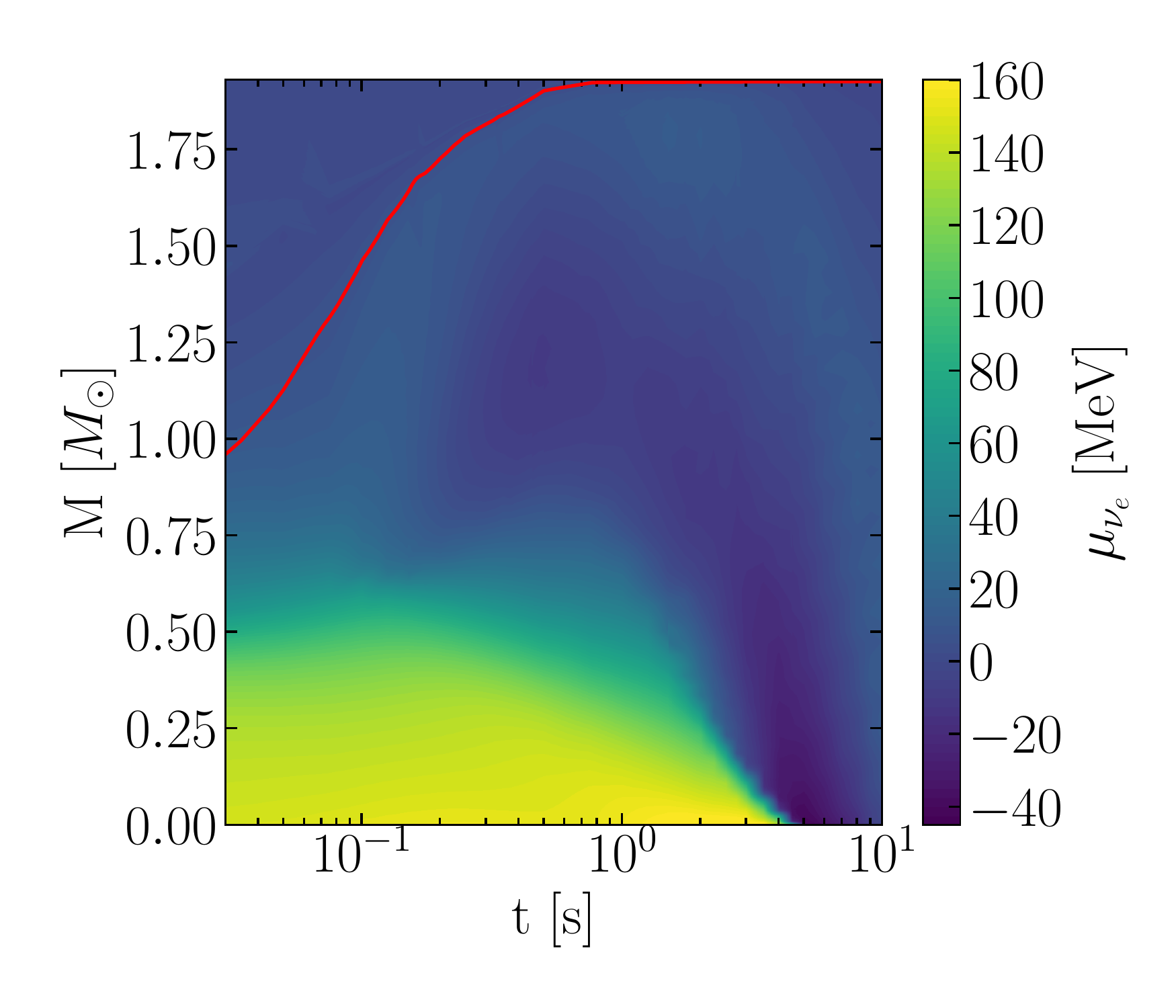}
    \includegraphics[width=0.32\textwidth,height=0.312\textwidth]{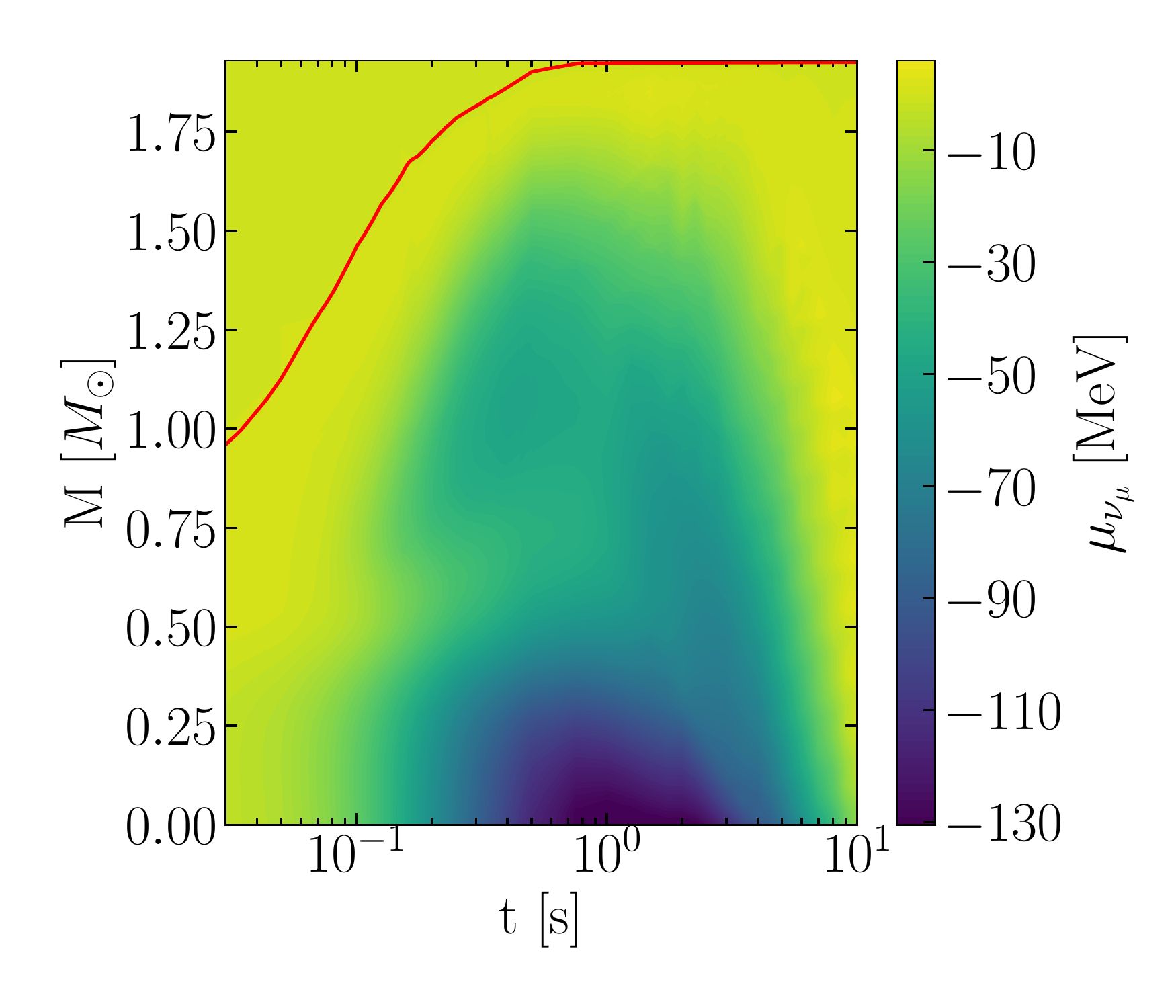}
    \caption{Same as Fig.~\ref{fig:coldmodelproperties} for the Garching ``hot'' model.
    The final neutron-star mass is here $1.926\,M_\odot$.
    \label{fig:hotmodelproperties}}
\vskip12pt
    \centering
    \includegraphics[width=0.36\textwidth]{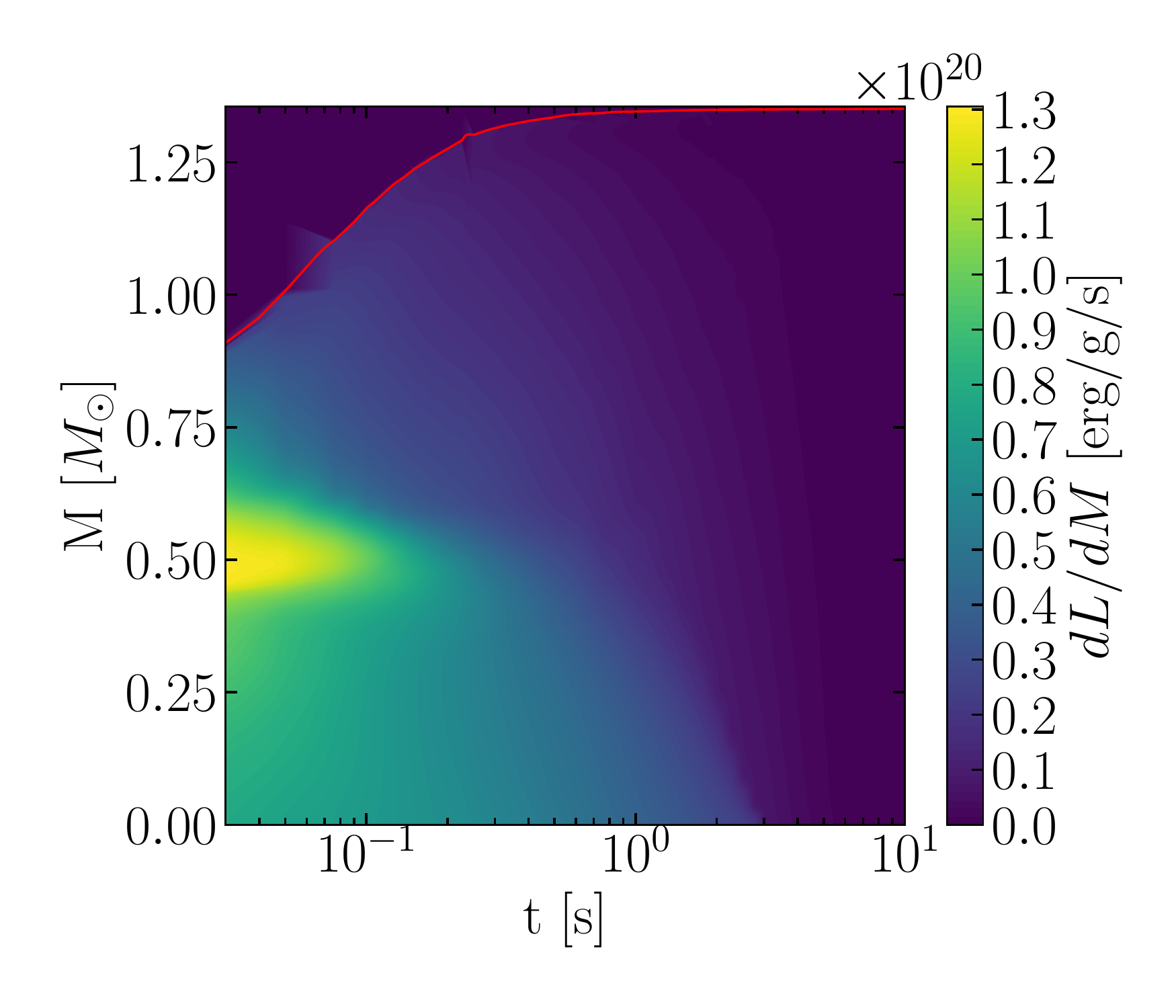}
    \includegraphics[width=0.36\textwidth]{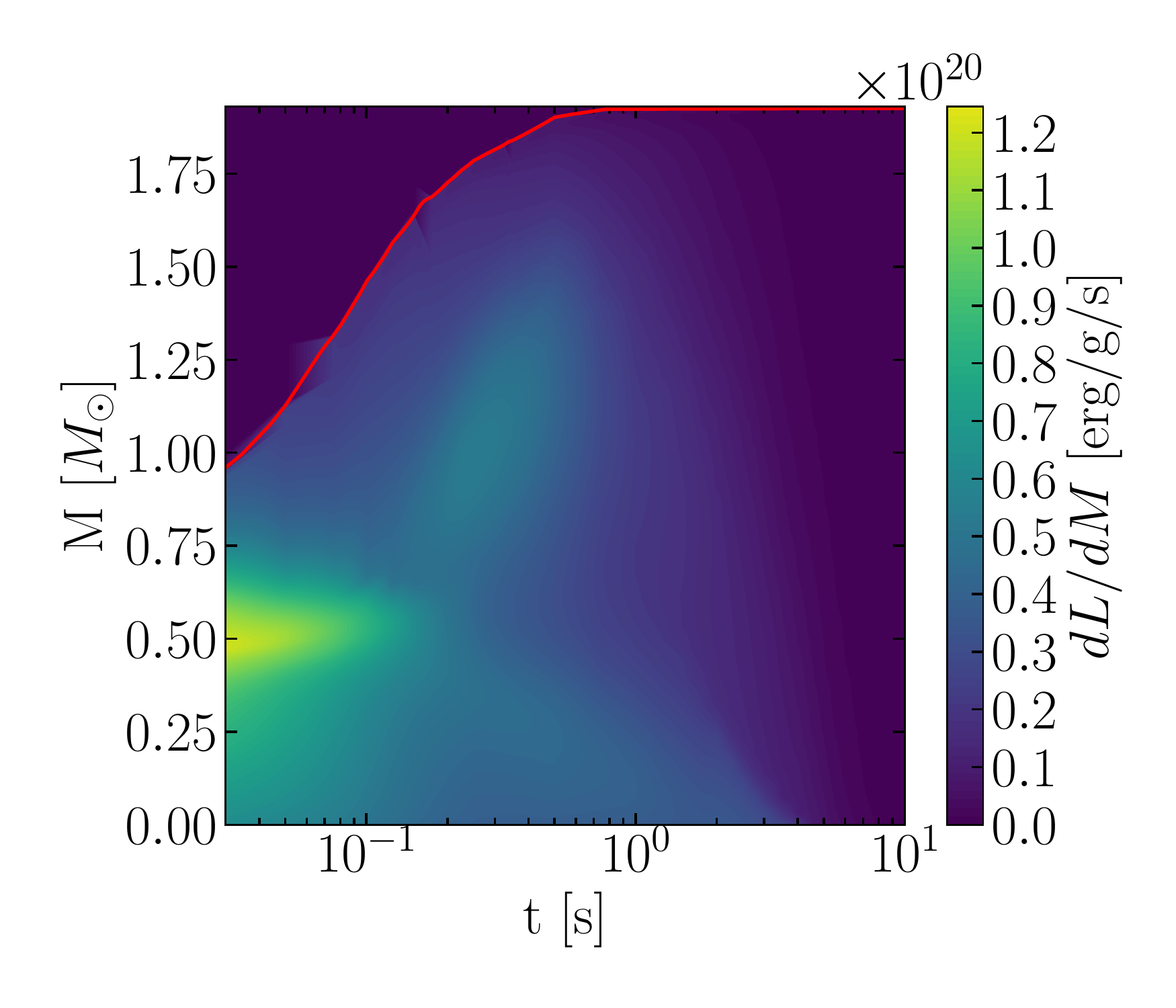}
    \caption{Majoron emissivity per unit mass as a function of time and mass coordinate, for the Garching ``cold'' (left, obtained using $g_\phi m_\phi= 8.3\times 10^{-9}$~ MeV) and ``hot'' (right, obtained using $g_\phi m_\phi= 7.7\times 10^{-9}$~ MeV) model. The red line identifies the density $3\times 10^{12}$ g cm$^{-3}$ and thus the edge of PNS as in the previous figures.}
    \label{fig:emissivitiesinplane}
\end{figure*}

In Figs.~\ref{fig:coldmodelproperties} and Figs.~\ref{fig:hotmodelproperties} we show several internal properties of these models as a function of time and mass coordinate for these two models. The left panels show the temperature and we see that after collapse the models are cold. They heat up at the edge of the inner core as they contract, with the maximum $T$ and largest extent of the hot region achieved at around 1~s. Therefore, the emission rate of new particles would be largest around this time if the emission rate depends on temperature, as it often happens in other extensions of the Standard Model because it is the thermal energy of the medium constituents that is emitted.

However, in our case of Majoron emission by neutrino coalescence, the process $\nu_e\nu_e\to\phi$ dominates by far, and so the chemical potential $\mu_{\nu_e}$ rather than $T$ is the key quantity. It is shown in the middle panels, and we see that it is some 100~MeV up to roughly the inner $0.5\,M_\odot$, corresponding roughly to a radius of 10~km. At 1--2~s it drops quickly as the core deleptonizes. Beta equilibrium implies that $\Delta\mu=\mu_e-\mu_{\nu_e}=\mu_\mu-\mu_{\nu_\mu}=\mu_n-\mu_p$, whereas the number densities of $\nu_e$ and $e^-$ must add up to the trapped lepton number of around 0.30 per baryon. However, the exact value of $\Delta\mu$ depends on the nucleon properties in the medium and thus on the equation of state. Using free protons and neutrons provides the right order of magnitude, but is not a good approximation to estimate the emission rate, because in our case the latter scales rapidly as $\mu_{\nu_e}^3$ (see main text).

In these models with six-species neutrino transport, a chemical potential also builds up for $\nu_\mu$ in the sense that a significant population of $\bar\nu_\mu$ builds up, but the maximum of $|\mu_{\nu_\mu}|$ remains a factor of 2--3 smaller than $\mu_{\nu_e}$. As the emission rate scales with $\mu_\nu^3$, the muonic contribution remains only an order 10\% correction.

In Fig.~\ref{fig:emissivitiesinplane}, we finally show contours of the Majoron emission rate per unit mass. While the Majoron emission rate per unit volume scales as $\mu_\nu^3$ and thus peaks at the center of the star, the emission rate per unit mass peaks at the edge of the inner core shown by the ``yellow peak''. This is because of the larger volume associated with the outer shells of the core. The chosen coupling strength for both ``hot'' and ``cold'' model coincides with the corresponding energy loss criterion detailed in the text, so that the Majoron luminosity at 1~s coincides with the neutrino luminosity. For the chosen coupling strength of $g_\phi m_\phi= 8.3\times 10^{-9}$~ MeV (cold) and $g_\phi m_\phi= 7.7\times 10^{-9}$~ MeV (hot), the emission rate is around $1\times10^{20}$~erg/gs throughout the inner core up to $0.50\,M_\odot$ for the first second and then drops quickly. In the hot model, there is significant emission at larger mass coordinate around 0.5~s, deriving from the relatively large $\bar\nu_\mu$ population.

\section{E.~Neutrino chemical potentials and older models}
\label{sec:mu-nu}

Previous authors have derived SN~1987A energy-loss bounds, based on the same coalescence process, or have provided sensitivity forecasts for 100-MeV-range events from a future galactic SN \cite{Heurtier:2016otg,Akita:2022etk}. The emitting SN core was approximated as a one-zone model with $\mu_{\nu_e}=200~{\rm MeV}$ over a volume with $R=10$~km and, in the case of Ref.~\cite{Akita:2022etk}, for a time scale of 10~s. These assumptions yield far more restrictive limits or far more ambitious signal predictions than our one-zone model or the numerical Garching models.

In Ref.~\cite{Akita:2022etk}, the chemical potential was taken from the pioneering paper \cite{Burrows:1986ApJ} (see their Fig.~11). In this proto-neutron star (PNS) cooling simulation, the nuclear equation of state was still relatively rough. Moreover, the starting value of trapped lepton number per baryon of $Y_L=0.35$ was chosen as an initial condition and did not follow from a self-consistent SN simulation. More recent systematic PNS cooling simulations \cite{Pons:1999ApJ} used more sophisticated nuclear and microphysics, chose a similar initial $Y_L=0.35$, and found an initial value at the center of $\mu_{\nu_e}\sim170$~MeV (see Fig.~9 for their baseline model).

Modern self-consistent simulations that include the infall phase find much smaller values of the trapped lepton number, 0.30 being a more typical number, depending on the progenitor model, also leading to smaller $\mu_{\nu_e}$. In the muonic Garching models used here, the trapped lepton number in the center at core bounce is around 0.28 for the hot and 0.29 for the cold model and the initial $\mu_{\nu_e}\sim150$~MeV at the center.

For Majoron emission, the geometrically largest region, very roughly around a mass coordinate of $0.5\,M_\odot$, is more relevant than the values at the center and so this region is indicative of the parameters that one could use for a one-zone description. This point is especially relevant for the time evolution because deleptonization occurs earlier at larger radii. Figure~11 of Ref.~\cite{Burrows:1986ApJ} reveals that after only a few seconds, $\mu_{\nu_e}$ strongly drops, and considering that the emission rate varies as $\mu_{\nu_e}^3$, the signal would strongly quench at 2--3~s and a similar conclusion follows from Fig.~9 of Ref.~\cite{Pons:1999ApJ}.

However, the deleptonization time scale can be much faster if the effect of PNS convection is included, in contrast to Refs.~\cite{Burrows:1986ApJ,Pons:1999ApJ} or recently Ref.~\cite{Li:2020ujl} who studied the late neutrino signal. We refer to a recent study of PNS evolution \cite{Pascal:2022MNRAS} (see this paper for references to the earlier literature) who found that convection, implemented with a mixing-length approximation, speeds up deleptonization by about a factor of~4 (see especially their Sec.~4.1) and as such is crucial for determining the overall time scale. Of course, the exact quantitative impact on Majoron emission or on SN neutrino signal properties may not be captured by this single number which refers to deleptonization at the center of the star.

In our study we have used the numerical Garching models described earlier that include a mixing-length treatment of PNS convection, use nuclear equations of state that agree with modern information (notably on neutron-star masses and radii), and find trapped lepton abundances and chemical potentials commensurate with other modern simulations.

For the case of Majoron emission one can actually characterize the different models with a single figure, the trapped number of $\nu_e$ in the core. In the degenerate limit, the Majoron luminosity of the SN core happens to be proportional to $N_{\nu_e}$, the total number of $\nu_e$ present in the core as explained around Eq.~\eqref{eq:Lphi} below. For our cold model at core bounce, we find $\hat{N}_\nu=3.5$ in units $(100~{\rm MeV})^3(10~{\rm km})^3$, whereas at 1~s postbounce it is 0.74. These numbers justify the one-zone parameters adopted in the main text. 

If instead one uses $\mu_\nu=200~{\rm MeV}$ with the same one-zone radius 10~km, at 1~s one finds $\hat{N}_\nu=8$, about a factor of 11 larger and thus leading to much more restrictive energy-loss bounds as reported, for example, in Refs.~\cite{Heurtier:2016otg,Akita:2022etk}.

For our argument about missing 100-MeV-range neutrinos in the SN~1987A data or the earlier forecasts for a future galactic SN \cite{Akita:2022etk}, what matters is a somewhat different quantity. In the degenerate limit, the number emission rate scales with $\mu_\nu^2$. If we assume very roughly that the detection cross section scales with energy-squared, the count rate arising from a one-zone model
scales with $\mu_\nu^4 R^3\tau$ as discussed in the main text. Therefore, one simple figure of merit for the source model is $\hat{C}_\nu=3\int dt\,dr\,r^2\mu_\nu^4(r,t)$. Our cold model yields \smash{$\hat{C}_\nu=2.30$} in units of
$(100~{\rm MeV})^4(10~{\rm km})^3\,{\rm s}$. If one were to use a one-zone model with $\mu_\nu=200~{\rm MeV}$, $R=10~{\rm km}$ and $\tau=10~{\rm s}$, one instead finds $\hat{C}_\nu=160$, a factor of 70 larger than our value.

The sensitivity of the $\nu_e$ abundance in the SN core, and its time-integrated value, to the microphysics input as well as deleptonization speed of the SN model mandates a somewhat careful gauging of one-zone parameters.

\section{F.~Majoron decay rate and emissivity}

The matrix element for the decay of a single Majoron into a pair of neutrinos is
\begin{equation}
|\mathcal{M}|^2=g^2 m_\phi^2.
\end{equation}
This is also the matrix element for coalescence of a pair of neutrinos into a Majoron; notice that there are no additional factors coming from averages over spin states since we consider Majorana neutrinos. 

The decay rate of a Majoron into a pair of neutrinos is
\begin{eqnarray}
\Gamma_{\phi\to\nu\nu}&=&\frac{1}{2}\int \frac{d^3\mathbf{p}_1}{(2\pi)^3 2E_1} \frac{d^3\mathbf{p}_2}{(2\pi)^3 2E_2}\frac{1}{2m_\phi}\\ \nonumber 
 &&{}\times(2\pi)^4\delta^{(4)}(p_1+p_2-p_\phi)|\mathcal{M}|^2,
\end{eqnarray}
where we denote by $p_1$, $p_2$, and $p_\phi$ the four-momenta of the two neutrinos and the Majoron respectively, and in bold we denote their three-momenta. The factor $1/2$ accounts for the presence of two identical particles in the final state. Performing the phase-space integral, we obtain
\begin{equation}\label{eq:vacuum-decay-rate}
\Gamma_{\phi\to\nu\nu}=\frac{g^2 m_\phi}{32\pi}.
\end{equation}

In the case of neutrino coalescence, the rate of Majoron production from a pair of neutrinos, restricting to a single flavor, is (see, e.g., Ref.~\cite{Raffelt:1996wa})
\begin{eqnarray}
\frac{d\dot{N}_\phi}{dE_\phi}&=&\frac{1}{2}\int \frac{d^3\mathbf{p}_1}{(2\pi)^3 2E_1} \frac{d^3\mathbf{p}_2}{(2\pi)^3 2E_2} \frac{|\mathbf{p}_\phi|}{4\pi^2}
\\ \nonumber
&&{}\times(2\pi)^4 \delta^{(4)}(p_1+p_2-p_\phi) f_\nu(E_1) f_\nu(E_2) |\mathcal{M}|^2,
\end{eqnarray}
where $f_\nu(E)$ is the neutrino phase-space distribution function. Performing the integral we recover
\begin{equation}\label{eq:BCE}
\frac{d\dot{N}_\phi}{dE_\phi}=\frac{g^2 m_\phi^2}{64\pi^3}\int_{E_-}^{E_+}dE_\nu f(E_\nu) f(E_\phi-E_\nu),
\end{equation}
as reported in the main text.

Actually it is instructive to compare the rate of absorption $\Gamma_{\rm A}(E_\phi)$ of a Majoron in the neutrino background with the spontaneous rate of emission $\Gamma_{\rm E}(E_\phi)$. The rate of absorption is given by the vacuum decay rate Eq.~\eqref{eq:vacuum-decay-rate} times a Lorentz factor $m_\phi/E_\phi$. Moreover, the final-state neutrinos are Pauli-blocked so that overall we find
\begin{equation}
    \Gamma_{\rm A}=\frac{g^2m_\phi^2}{32\pi\,E_\phi}
    \int_{E_-}^{E_+}\frac{dE_\nu}{p_\phi}  [1-f(E_\nu)][1-f(E_\phi-E_\nu)],
\end{equation}
where $E_{\pm}=\frac{1}{2}(E_\phi\pm p_\phi)$ as defined in the main text. The integral expression is equal to 1 in the absence of Pauli blocking because the interval of integration has length~$p_\phi$.

On the other hand, the emission rate per unit volume given in Eq.~\eqref{eq:BCE} is equal to
\begin{equation}
    \frac{d\dot{N}_\phi}{dE_\phi}=\Gamma_{\rm E}(E_\phi)\,\frac{4\pi p_\phi^2}{(2\pi)^3}
    \,\frac{E_\phi}{p_\phi}
\end{equation}
in terms of the spontaneous emission rate $\Gamma_{\rm E}$ and a Majoron phase-space factor. The last factor is the Jacobian from changing a $dp_\phi$ to a $dE_\phi$ integration. Therefore, we find
\begin{equation}
    \Gamma_{\rm E}(E_\phi)=\frac{g^2m_\phi^2}{32\pi\,E_\phi}
    \int_{E_-}^{E_+}\frac{dE_\nu}{p_\phi}  f(E_\nu)\,f(E_\phi-E_\nu)
\end{equation}
for the spontaneous emission rate.

The neutrinos follow a Fermi-Dirac distribution at temperature $T$ and chemical potential $\mu$ so that $f(E_\nu)=(e^{(E_\nu-\mu)/T}+1)^{-1}$. Explicit integration reveals that
\begin{equation}\label{eq:detailed-balance}
    \frac{\Gamma_{\rm E}(E_\phi)}{\Gamma_{\rm A}(E_\phi)}=\exp\left(-\frac{E_\phi-2\mu}{T}\right).
\end{equation}
If the decay were of the form $\phi\to\nu\bar\nu$, the Pauli-blocking and occupation-number factors would involve a FD distribution with $+\mu$ and one with $-\mu$ and then we would find the usual detailed-balance factor $e^{-E_\phi/T}$. However, in our case the neutrino medium is not in equilibrium because Majoron emission destroys neutrino pairs and reduces the chemical potential. Equilibrium here would mean vanishing $\mu$. This explains the unusual detailed-balance factor of Eq.~\eqref{eq:detailed-balance}.

If the neutrinos are perfectly degenerate, the occupation number is $f(E_\nu)=\Theta(\mu_{\nu}-E_\nu)$. If the Majoron mass is small compared with the energies, the integral in Eq.~\eqref{eq:BCE} is a triangle function that linearly rises from 0 to $\mu_\nu$ for $0\leq E_\phi<\mu_\nu$ and then linearly decreases to zero at $E_\phi=2\mu_\nu$. Integrating Eq.~\eqref{eq:BCE} over $\int dE_\phi\,E_\phi$
yields the energy-loss rate per unit volume of
\begin{equation}
    Q_\phi=\frac{g^2 m_\phi^2}{64\pi^3}\,\mu_\nu^3,
\end{equation}
as reported in the main text. In this case, the Majoron luminosity of the SN core is
\begin{equation}\label{eq:Lphi}
    L_\phi=\frac{g^2 m_\phi^2}{64\pi^3} \frac{4\pi}{3} \hat{N}_\nu
    \quad\hbox{with}\quad
    \hat{N}_\nu=3\int_0^\infty dr\,r^2\,\mu_\nu(r)^3.
\end{equation}
In a one-zone model with the radius $R$ and constant chemical potential $\mu_\nu$, we thus have
$\hat{N}_\nu=(\mu_\nu R)^3$, which is a dimensionless number. The density of degenerate neutrinos is $n_\nu=(4\pi/3)\,(\mu_\nu/2\pi)^3=\mu_\nu^3/6\pi^2$ whereas the spatial volume is $V=(4\pi/3)R^3$, so 
$\hat{N}_\nu=(2/9\pi)\,n_\nu V$ and up to a numerical factor the total number of trapped neutrinos.

\end{document}